\documentclass[amsmath,amssymb,superscriptaddress,notitlepage]{revtex4-1}

\usepackage{graphicx,amsmath,amssymb,amstext,amsfonts,color,dsfont,epstopdf,mathtools,bbm,geometry}
\usepackage[colorlinks,linkcolor=black,citecolor=blue,urlcolor=blue]{hyperref}
\usepackage[latin9]{inputenc} % For Schr_ö_dinger etc.
\usepackage{times}
\mathtoolsset{centercolon}
\newcommand{\one}{\mathbbm{1}}
\DeclareMathOperator{\id1}{\mathds{1}}

\newcommand{\pTr}[2]{\mathrm{Tr}_{\mathrm{#1}}\left[ {#2} \right]}
\newcommand{\tr}{{\mathrm{Tr}}}
\newcommand{\be}{\begin{equation}}
\newcommand{\ee}{\end{equation}}
\newcommand{\ben}{\begin{eqnarray}}
\newcommand{\een}{\end{eqnarray}}
\newcommand\dx{\mathrm{d} x}

\newcommand\cc{\mathbb{C}}

\newcommand\im{\mathrm{i}}

\newcommand{\ket}[1]{|#1\rangle}
\newcommand{\bra}[1]{\langle#1|}

\newcommand{\CR}[1]{\textcolor{black}{#1}}
\newcommand{\AS}[1]{\textcolor{black}{#1}}

\DeclareMathOperator{\po}{\mathcal{P}}
\DeclareMathOperator{\ps}{\hat{\Psi}}
\DeclareMathOperator{\psd}{\hat{\Psi}^{\dagger}}

\definecolor{jens}{rgb}{0,.8,.5}

\begin{document}

\title{Quantum field tomography}

\author{A. Steffens, C. A. Riofr\'io, R. H{\"u}bener, and J.\ Eisert}
\affiliation{Dahlem Center for Complex Quantum Systems, Freie Universit\"{a}t Berlin, 14195 Berlin, Germany}

\date{\today}

\begin{abstract}
We introduce the concept of quantum field tomography, the efficient and reliable reconstruction of unknown quantum fields based on data of correlation functions. At the basis of the analysis is the concept of continuous matrix product states, a complete set of variational states grasping states in \CR{one-dimensional} quantum field theory. We innovate a practical method, making use of and developing tools in estimation theory 
used in the context of compressed sensing such as Prony methods and matrix pencils, allowing us to faithfully reconstruct quantum field states based on low-order correlation functions. In the absence of a phase reference, we highlight how  specific higher order correlation functions can still be predicted. We exemplify the functioning of the approach by reconstructing randomised continuous matrix product states from their correlation data and study the robustness of the reconstruction for different noise models. Furthermore, we apply the method to data generated by simulations based on continuous matrix product states and using
the time-dependent variational principle. The presented approach is expected to open up a new window into  experimentally studying continuous quantum systems, such as encountered in experiments with ultra-cold atoms on top of atom chips. By virtue of the analogy with the input-output formalism in quantum optics, it also allows for studying open quantum systems.

\end{abstract}

\maketitle
%\pacs{05.45.Mt, 03.65.Wj, 42.50.Lc, 42.50.Ct, 03.65.Yz}

\tableofcontents

\section{Introduction}
Quantum theory predicts probability distributions of outcomes in anticipated quantum measurements. The 
actual problem encountered in practice, however, is often not so much concerned with predicting certain outcomes of specific measurement procedures, but rather with reconstructing the unknown quantum state at hand that is compatible with precisely such  measurement outcomes. This task of reconstructing states based on data---possible under certain conditions of 
completeness or other reasonable assumptions---is called {\it quantum state tomography}. For finite-dimensional quantum systems, this task is feasible and is routinely used in experiments.
However, the number of parameters to be determined scales exponentially with the system size: Full quantum state tomography is highly inefficient. This is even so much less of a problem than one might at first be tempted to think. It was one of the major insights in the field in recent years to recognise that economical or efficient quantum state tomography is well possible for systems with many degrees of freedom. In fact, in most physically relevant questions, fully unconstrained quantum state tomography may be said to solve the ``wrong problem''. One is surely often not interested in arbitrary states, but only in those states that one is expected to encounter in practice, which are naturally more restricted.%} 

\AS{In the context of {\it compressed sensing tomography} \cite{Compressed,Compressed2} or {\it matrix product states tomography} \cite{MPSTomo,MPOTomo,Efficient}, identification of quantum systems with many degrees of freedom is indeed well possible. The key step is to identify} the right model in which to represent the states, e.g., approximately low-rank states or those with clustering correlation functions. In the context of matrix product state tomography, the notion of a model refers to a meaningful variational class of states that provably captures all states exhibiting low entanglement \cite{EisertAreaLaws,SchuchApprox}.  In this sense, 
tomography is efficiently possible for any system size.  In fact,  by increasing the bond dimension, an arbitrary state can be well approximated. Quite similar to the mindset of
compressed sensing, a ``sparsity of commonly encountered states'' is heavily used for the benefit of tomography.

In quantum field theory, \AS{where one has to consider an infinite number of degrees of freedom, the situation is in principle aggravated}. Analogously, a moment of thought reveals that to think about quantum field tomography in the sense of trying to ``fill an infinite table with numbers'' is rather ill-guided. This is not the actual problem one aims at solving in any practical context---one again needs to identify the appropriate model and the right ``sparsity structure''.

In this work, we introduce the concept of {\it quantum field tomography}, tomography of continuous systems in quantum field theory, and provide a practical and feasible method for achieving this. We do so by drawing and further developing ideas from the study of {\it continuous matrix product  states} \cite{PhysRevLett.104.190405,Osborne2010,cMPS_Calc}, methods of how to assess higher order correlation functions in that context \cite{Wick_MPS}, as well as a machinery from statistical estimation theory, such as a {\it Prony analysis} \cite{Prony1795} and {\it matrix pencil methods} \cite{HuaSarkarGPOF,HuaSarkarMPM1990}, which are here brought to a new context. In fact, these methods of estimation have not been considered before in the context of quantum state reconstruction and are expected to be interesting in their own right. The basis of the analysis are low-order multi-point correlation functions directly accessible in many common current experiments.

This approach opens up a new window into grasping the physics of continuous quantum systems in equilibrium and non-equilibrium. Instead of having to make a physical model (e.g., define a Hamiltonian) and checking for the plausibility of it, one can---based on data of correlation functions---reconstruct the quantum field itself. Such an approach seems particularly appealing when studying one-dimensional continuous bosonic models such as {\it ultra-cold atoms on top of atom chips} \cite{SchmiedmayerScience,KitawaSchmiedmayer2011NJPh,LangenNature}. What is more, if only partial data is available, say, in the absence of a phase reference frame, higher-order correlation functions of the same type can be predicted as well. The starting point of the analysis is what is called ``Wick's theorem for matrix product states'' \cite{Wick_MPS}, which is here brought to a new level and transformed into a practical method of reconstructing unknown continuous matrix product states from correlation function data.
%}

This work is structured as follows. In Sec.\ \ref{sec:background}, we will give a short overview of the concept of
continuous matrix product states (cMPS) \cite{PhysRevLett.104.190405,Osborne2010,cMPS_Calc} as well as what can be called a ``Wick theorem'' for this class of states \cite{Wick_MPS}, aiming as a preparation for the following technical sections. In Sec.\ \ref{sec:reconstruction}, we will describe in great technical detail how to reconstruct a field state from its low order correlation functions and give a complete matrix product state description of it. The limitations of this method are investigated in Sec.\ \ref{sec:limitations}. In Sec.\ \ref{sec:applications}, we will demonstrate the method using simulated data from random cMPS and apply the method to the ground state of the Lieb-Liniger model, a prototypical integrable model in quantum
field theory \cite{1963PhRv..130.1605L,Caux}. 
The data used here have been generated using a cMPS-based simulation based on the time-dependent variational principle \cite{2011PhRvL.107g0601H,2013PhRvL.111b0402D,cmps_groundstate}.
The impact of noise in real world-scenarios on the method is investigated here. In Sec.\ \ref{sec:summary}, we summarise and conclude this work.

%%%%%%%%%%%%%%%%%%%%%%%%%%%%%%%%%%%%%%%%%%%%%%%%%%%%%%%%%%%%%%%%%%%%%%%%%%%%%%
%%%%%%%%%%%%%%%%%%%%%%%%%%%%%%%%%%%%%%%%%%%%%%%%%%%%%%%%%%%%%%%%%%%%%%%%%%%%%%
%%%%%%%%%%%%%%%%%%%%%%%%%%%%%%%%%%%%%%%%%%%%%%%%%%%%%%%%%%%%%%%%%%%%%%%%%%%%%%

\section{Background}\label{sec:background}

In this work, we are concerned with one-dimensional quantum fields with fast decaying spatial correlations. Analogous to the case of many-body quantum systems, successfully described by the 
{\it matrix product state} (MPS) formalism, there is a variational class of 
states specially suited to study such systems: the {\it continuous matrix product states} (cMPS) \cite{PhysRevLett.104.190405,Osborne2010}. 

\subsection{Continuous matrix product states}

In this section, we briefly review the basics of the cMPS formalism. For a review and comprehensive discussion of the computation of correlation functions,  see, e.g., Ref.\ \cite{cMPS_Calc}.

\subsubsection{Basic definitions}
A translationally invariant cMPS with periodic boundary conditions and one species of bosonic particles is defined as
\begin{equation}
\ket{\psi_{Q,R}}=\pTr{aux}{\po\textrm{e}^{\int_{0}^{L}\dx \left(Q\otimes\hat{\one}+R\otimes\hat{\Psi}^{\dagger}(x)\right)}}\ket{\Omega},\label{eq:cMPS:def}
\end{equation}
where the collection of field operators $\hat\Psi(x)$, $x\in[0,L]$, 
obey the bosonic commutation relations of the free field  
\begin{equation}
	[\text{\ensuremath{\hat{\Psi}}}(x), \text{\ensuremath{\hat{\Psi}}}^{\dagger}(y)] =\delta(x-y), 
\end{equation}
$\left|\Omega\right\rangle$ is the vacuum state vector, 
$Q,R\in\cc^{d\times d}$ are matrices acting on an {\it auxiliary $d$-dimensional space} $\mathcal{A}$, the ``virtual space'', and constitute the variational parameters of the class. $L$ is the length of the closed physical system, $\mathcal{P}$ denotes the path ordering operator and $\mathrm{Tr_{aux}}$ traces out the auxiliary space.

The parametrisation in \eqref{eq:cMPS:def} by $Q$ and $R$ is not unique, i.e., there is and additional \emph{gauge freedom}. Namely, when simultaneously conjugating $Q$ and $R$ 
with an invertible matrix $G$  \cite{cMPS_Calc},
\begin{align}
\tilde{Q} & =G^{-1}QG ,\\
\tilde{R} & =G^{-1}RG,\label{eq:cMPS:Gauge formula}
\end{align}
then the two resulting state vectors still represent the same state, i.e., all expectation values are invariant under this transformation. 

\subsubsection{Related physical processes}\label{sec:Lindblad}

A useful interpretation of the correlations in cMPS can be given in terms of a $d$-dimensional (auxiliary) quantum system $\mathcal{A}\cong\mathbb{C}^d$ interacting with a \AS{one-dimensional }field~$\mathcal{F}$~\cite{Osborne2010}. The Hamiltonian of the joint system 
is given by  
\begin{equation} 
	\hat{H}\left(x\right)=K\otimes\hat{\one}_{\mathcal{F}}+R\otimes\hat{\Psi}^{\dagger}(x)+R^{\dagger}\otimes\hat{\Psi}(x),
\end{equation}
where $\hat{\one}_{\mathcal{F}}$ is the identity on the field, $K\in\cc^{d\times d}$  the Hamiltonian of the free evolution of the finite dimensional system, and $R\otimes\hat{\Psi}^{\dagger}(x)$ the coupling between the system and the field with $R\in\cc^{d\times d}$. \AS{Note that $H$ evolves in position, rather than time---in this picture, both are by construction equivalent. }Starting with the state vector
$\ket{\varphi_{i}}\ket{\Omega}$, where $\ket{\varphi_{i}}\in\mathcal{A}$ and the vacuum $\ket{\Omega}\in\mathcal{F}$, and evolving over $\left[0,L\right] \ni x$, we formally arrive at 
\begin{equation}
\hat{U}\left(0,L\right)\ket{\varphi_{i}}\ket{\Omega}:=\po\textrm{e}^{-\mathrm{i}\int_{0}^{L}\dx\left(K\otimes\hat{\one}_{\mathcal{F}}-\frac{1}{2}R^{\dagger}R\otimes\hat{\one}_{\mathcal{F}}+\mathrm{i}R\otimes\hat \Psi^{\dagger}(x)\right)}\ket{\varphi_{i}}\ket{\Omega},
\end{equation}
\AS{using the Baker-Campbell-Hausdorff formula and the fact that $R^{\dagger}\otimes\hat{\Psi}(x)\ket{\varphi_{i}}\ket{\Omega}=0$.} By setting 
\begin{equation}\label{eq:KQ}
	Q=-\textrm{i}K- \frac{1}{2}R^{\dagger} R,
\end{equation}
projecting onto $\bra{\varphi_{i}}\otimes\hat{\one}_{\mathcal{F}}$ to decouple $\mathcal{A}$ from $\mathcal{F}$, and summing over a complete orthonormal basis of all $\ket{\varphi_{i}}$, we again obtain Eq.~\eqref{eq:cMPS:def}. This shows the interpretation of the cMPS formalism in the sequential preparation picture of MPS \cite{MPSSurvey}.

In this picture, we interpret $K$ to be the Hamiltonian of a virtual particle in the auxiliary space that mediates field interactions. Even more \cite{Osborne2010}, the dynamical 
behaviour of the auxiliary system $\mathcal{A}$ can be modelled by computing the derivative of \begin{equation}
\rho_{\mathcal{A}}\left(x\right)=\pTr{\mathcal{F}}{\hat{U}\left(x,L\right)\left(\rho_{\mathcal{A}}\left(0\right)\otimes\ket{\Omega}\bra{\Omega}\right)\hat{U}^{\dagger}\left(x,L\right)},
\end{equation}
where $\tr_{\mathcal{F}}{}$ means tracing out the physical system $\mathcal{F}$. This yields the ordinary differential equation \begin{equation}
\frac{\mathrm{d}}{\dx}\rho_{\mathcal{A}}(x)=-\mathrm{i}\left
[K,\rho\left(x\right)\right]+R^{\dagger}\rho(x)R-\frac{1}{2}\left[R^{\dagger}R,\rho\left(x\right)\right]_{+},
\end{equation}
which is a master equation in Lindblad form, governing the Markovian evolution of $\rho_{\mathcal{A}}$, where $R$ plays the role of dissipative quantum jump (Lindblad) operators. 
Although arbitrary $Q$ and $R$ lead to a valid cMPS, not all pairs give rise to an effective Hamiltonian $K$ via eq.~\eqref{eq:KQ}. For this, it is required that 
\begin{equation}	
	Q+Q^{\dagger}+R^{\dagger}R=0.
\end{equation}
However, arbitrary $Q$ and $R$ can in general be transformed into a specific gauge where they fulfil this equation.

\subsection{Correlation functions in cMPS}

The mathematical relations between the $n$-point functions are the starting point for our tomography algorithms, hence we give a brief summary at this point. A quantum field state can be completely characterised by all the possible normal expectation values constructed from $\hat{\Psi}(.)$ and $\hat{\Psi}^\dag(.)$ and their commutation relations. In this work, 
we will focus on \emph{density-like} correlation functions, i.e., for each position $x_k\in[0,L]$, $k=1,\dots,n$, both operators $\hat{\Psi}^\dag(x_k)$ and $\hat{\Psi}(x_k)$ exist within the expectation values. Because of translational invariance, we can set $x_1=0$ without loss of generality.
The expectation value $\langle\psi_{Q,R}|\hat{\Psi}^{\dagger}(x_{1})\dots\hat{\Psi}^{\dagger}(x_{n})\hat{\Psi}(x_{n})\dots\hat{\Psi}^{\dagger}(x_{1})|\psi_{Q,R}\rangle$
can be computed as 
%and setting x_1=0
\begin{equation}
	C^{(n)}(\tau_{1},\dots,\tau_{n-1}):=\tr\left[\textrm{e}^{T\tau_{n}}\left(\overline{R}\otimes R\right)\dots\textrm{e}^{T\tau_{2}}\left(\overline{R}\otimes R\right)\textrm{e}^{T\tau_{1}}
	\left(\overline{R}\otimes R\right)\right],\label{eq:expval1}
\end{equation}
(see, e.g., Ref.\ \cite{cMPS_Calc}),
with the transfer matrix 
\begin{equation}	
	T:=\overline{Q}\otimes\id1_d+\id1_d\otimes Q+\overline{R}\otimes R,
\end{equation}
and the positive distances $\tau_{j}=x_{j+1}-x_{j}$ for $j= 1,\dots,n-1$ and $\tau_{n}=L-x_{n}$; the overline denotes complex conjugation. Correlation functions of cMPS are given by expressions involving only the auxiliary space. Static properties of a quantum field with one spatial dimension are hence related to non-equilibrium properties of a zero-dimensional system.
In this sense, they have been referred to as being ``holographic quantum states'' \cite{Osborne2010}.

For a normalised cMPS, the eigenvalues of $T$ are all complex with negative or zero real parts, due to the analogy to quantum channels \cite{Wolf2010}. %\AS{(src)} 
This leads to finite expectation values in the thermodynamic limit $L\rightarrow\infty$. Furthermore, assuming that $T$ is diagonalisable, which is in particular the case if its spectrum is non-degenerate, the $n$-point function \eqref{eq:expval1} can be further simplified to a sum of exponentially damped oscillatory terms
\begin{equation}
\lim_{L\rightarrow\infty} C^{(n)}(\tau_{1},\dots,\tau_{n-1})=\sum_{k_1,\dots, k_{n-1}=1}^{d^{2}}\rho_{k_{1},k_{2},\dots,k_{n-1}}\textrm{e}^{\lambda_{k_{1}}\tau_{1}}\dots\textrm{e}^{\lambda_{k_{n-1}}\tau_{n-1}}\label{eq:expval}
\end{equation} 
where 
\begin{equation}\label{eq:rho_M}
\rho_{k_{1},k_{2},\dots,k_{n-1}}=M_{1,k_{n-1}}M_{k_{n-1},k_{n-2}}\dots M_{k_{1},1}.
\end{equation} 
 The matrix $M\in\cc^{d^2\times d^2}$ is defined as $M=X^{-1}\left( \overline{R}\otimes R\right) X$, where $X$ is a change-of-basis matrix such that $X^{-1}TX$ is diagonal and compatible with the ordering of the eigenvalues $\{\lambda_k\}$. In the following, we will work exclusively in the thermodynamic limit and, for simplicity, use $C^{(n)}$ also to denote $n$-point correlation functions in this limit.

A first step to reconstruct a cMPS would be to identify $\{\rho_{k_{1},k_{2},\dots,k_{n-1}}\}$ and $\{\lambda_k\}$. That this is in principle possible can be seen by considering the Laplace transform of $C^{(n)}$
\begin{equation}\label{laplacedef}
\mathcal{L}^{(n)}(\mathbf{s})=\int_0^\infty d^{n-1} \boldsymbol{\tau} e^{-\mathbf{s} \cdot \boldsymbol{\tau}} C^{(n)}(\boldsymbol{\tau}),\quad s_1,\dots, s_{n-1} \in \mathbbm{C},
\end{equation}
which has the simple form
\begin{equation}\label{laplace2}
\mathcal{L}^{(n)}(\mathbf{s})=\sum_{k_1,\dots, k_{n-1}=1}^{d^{2}}\frac{\rho_{k_{1},k_{2},\dots,k_{n-1}}}{(\lambda_{k_{1}}-s_{1})\cdots(\lambda_{k_{n-1}}-s_{n-1})} .
\end{equation}
Each of the $d^{2(n-1)}$ combinations of $T$ eigenvalues appears as a pole of $\mathcal{L}^{(n)}$ in $\cc^{n-1}$ together with the corresponding residue in the numerator. If all the eigenvalues are different, i.e., the spectrum of $T$ non-degenerate, and all residues non-zero, then all residues are distinguishable as well. Since the Laplace transform itself proved to be infeasible for practical reconstruction algorithms, we will present alternative ways in the following. Independently of this, we want to keep calling the eigenvalues $\{\lambda_k\}$ the poles and $\{\rho_{k_{1},k_{2},\dots,k_{n-1}}\}$ the residues of the $n$-point function.
In the following, we require the spectrum of $T$ to be non-degenerate.

The structure of the correlation functions with the residues as products of entries of one matrix, Eq. (\ref{eq:rho_M}), allows for expressing higher order correlation functions by lower order correlation functions, very much reminding of the Wick's theorem in quantum field theory \cite{Wick_MPS}. In this sense, we will recover $M$ from the residues. We will describe this in detail below. 

\subsection{Additional symmetries}\label{sec:add_sym}

In the remainder of this work, we will make use of some symmetries that the cMPS fulfil. Here, we briefly state them. By construction, for each non-real entry of $\overline{R}\otimes R$ and $T$ there exists another entry containing its complex conjugate. 
More precisely, one can show that 
\begin{equation}
	\Lambda_d\overline{\overline{R}\otimes R}\Lambda_d=\overline{R}\otimes R
 \end{equation}
 and 
 $\Lambda_d\overline{T}\Lambda_d=T $,
 with  
 \begin{equation}\Lambda_d:=\sum_{j,k=1}^{d}E_{j,k}\otimes E_{k,j}
 \end{equation}
 and $E_{j,k}=e_{j} e_{k}^{T}$, the dyadic product of the canonical column vectors $e_j$, \cite[Sec.\ 2.5]{Graham_Kron}. Hence, if $\lambda$ is an eigenvalue of $T$ with eigenvector $v$ then $\Lambda_d\overline{T}\Lambda_dv=\lambda v$, and since $(\Lambda_d)^2=\one_{d^2}$, we obtain $T(\Lambda_d\overline{v})=\overline{\lambda}(\Lambda_d\overline{v})$, such that the spectrum of $T$ is closed under complex conjugation. This fact also follows from the channel property of cMPS as discussed in Ref.\ \cite{Wolf2010}.

For the reconstruction algorithms we will discuss below, it is instrumental to fix an unambiguous ordering of the eigenvalues of the transfer matrix $T$, which makes its diagonal matrix $D$ and furthermore the matrix $M$ unambiguous, too. If we order the eigenvalues in $D$ such that the $\kappa\in\{1,\dots ,d^2\}$ real eigenvalues constitute a block and the remaining $d^2-\kappa$ are arranged in complex conjugate pairs (e.g., ordering by descending real part), then $D$ obeys the symmetry relation $\Xi_{d,\kappa} \overline{D} \Xi_{d,\kappa}=D$ with the permutation matrix 
\begin{equation}
	\Xi_{d,\kappa}:=\one_{\kappa}\oplus\left(\bigoplus_{j=1}^{(d^{2}-\kappa)/2}\sigma_{x}\right)
\end{equation}
where $\sigma_x$ is the $x-$Pauli matrix. In addition, since $X$ consists of the eigenvectors $v$ of $T$ as column vectors, $\Lambda_d\overline{v}$ is the eigenvector of $\overline{\lambda}$, when $v$ corresponds to $\lambda$. Moreover, since $\Xi_{d,\kappa}$ interchanges the columns back, we have that $\Lambda_{d} \overline{X} \Xi_{d,\kappa}=X$. Using this fact and the definition $M=X^{-1}\overline{R}\otimes RX$, we obtain the symmetry relation $\Xi_{d,\kappa}\overline{M}\Xi_{d,\kappa}=M$ for the matrix $M$. This relation connects each  entry of $M$ with its complex conjugate and, via Eq. \eqref{eq:rho_M}, each residue with its complex conjugate. As with the poles, the set of residues is closed under complex conjugation for density-like correlation functions. These symmetries can also be used for a systematic least squares approach to reconstruct the poles and residues, see Sec.\ \ref{sub:recon_poles_res}.

%%%%%%%%%%%%%%%%%%%%%%%%%%%%%%%%%%%%%%%%%%%%%%%%%%%%%%%%%%%%%%%%%%%%%%%%%%%%%%
%%%%%%%%%%%%%%%%%%%%%%%%%%%%%%%%%%%%%%%%%%%%%%%%%%%%%%%%%%%%%%%%%%%%%%%%%%%%%%
%%%%%%%%%%%%%%%%%%%%%%%%%%%%%%%%%%%%%%%%%%%%%%%%%%%%%%%%%%%%%%%%%%%%%%%%%%%%%%

\section{State reconstruction}\label{sec:reconstruction}

Having established the structure of the correlation functions in cMPS, i.e., the structure of the data of our reconstruction problem, it remains to develop an appropriate protocol to extract the information encoded in the data. Given an $n$-point density-like correlation function of order 3 or higher corresponding to a cMPS $\ket{\Psi_{Q,R}}$, we will show that, in most cases, it is in principle possible to reconstruct the parameter matrices $Q$ and $R$ up to an arbitrary gauge and phase, and to reproduce all $n$-point functions.

We are dealing with a so-called inverse problem, a large class of
problems that make ``use of the actual results of some measurements of the observable parameters to infer the
actual values of the model parameters" \cite{tarantola2005inverse}. Many inverse problems are ill-conditioned---a small change in the
measurements can lead to a huge change in the model parameters. In this chapter we will examine the required
steps for cMPS reconstruction, see Fig.~\ref{fig:Reconstruction-steps}, and the respective main factors that influence their performance
regarding perturbed input data. Each step will be discussed in a separate section. We will see that in particular
the first and the last step can be notably ill-conditioned.

\subsection{Reconstruction steps}\label{sec:steps}

\begin{figure}
\includegraphics[width=1\columnwidth]{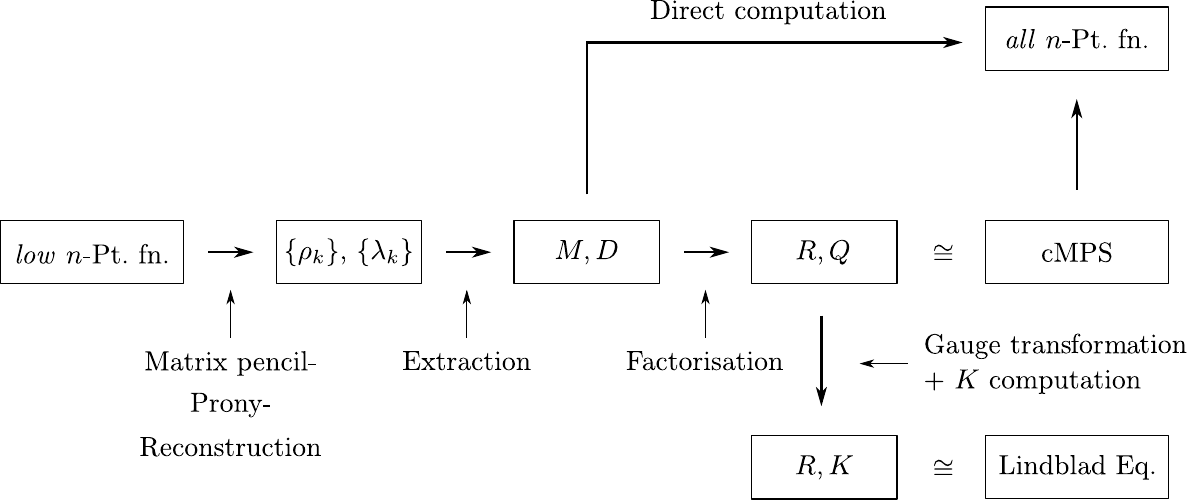}
\caption{\label{fig:Reconstruction-steps}The particular reconstruction steps starting with the input data, an $n$-point correlation function of a cMPS $\ket{\Psi_{Q,R}}$, and ending with the variational parameter matrices $Q$ and $R$, that fully characterise the state. Alternatively, the state can likewise be described by $K$ and $R$. With this knowledge, one can compute other $n'$-point correlation functions and compare with the input data to obtain evidence for a successful reconstruction.} 
\end{figure}
The reconstruction of a generic, translationally invariant cMPS in the thermodynamic limit comprises the following steps, which are represented in Fig.\ \ref{fig:Reconstruction-steps}:
\begin{enumerate}
\item The first step in processing the input data is to extract the poles $\left\{ \lambda_{k}\right\} $ and the residues $\{ \rho_{k_{1},k_{2},\dots,k_{n-1}}\} $ from a density-like $n$-point correlation function, $n\ge 3$,
\begin{equation}
C^{(n)}(\tau_{1},\dots,\tau_{n-1})=\sum_{k_1,\dots, k_{n-1}=1}^{d^{2}}\rho_{k_{1},k_{2},\dots,k_{n-1}}\textrm{e}^{\lambda_{k_{1}}\tau_{1}}\dots\textrm{e}^{\lambda_{k_{n-1}}\tau_{n-1}},
\end{equation}
which is measured and contains additional noise and experimental imperfections. 

\item In the second step, the matrix $M$ is determined from the residues
\begin{equation}
\rho_{k_{1},k_{2},\dots,k_{n-1}}=M_{1,k_{n-1}}M_{k_{n-1},k_{n-2}}\dots M_{k_{1},1},
\end{equation} 
and the matrix $D$ is determined from the poles. This can be achieved using certain invariances in the correlation functions that led to the formulation of Wick's theorem for matrix product states.
\item In the final step, the cMPS parametrisation matrices $Q$ and $R$ can be extracted from the matrices $M$ and $D$ by imposing a specific gauge. Additionally,  and after another gauge transformation, the Hamiltonian $K$ of the auxiliary system can be computed from the matrices $Q$ and $R$.
\end{enumerate}

In order to only generate and predict higher order density-like $n$-point functions, it is in general sufficient to use
the matrices $D$ and $M$ from the second step without any further reconstruction steps. This is in general much more robust against noise than the full reconstruction. Furthermore, we can leave out some of the poles (together
with the corresponding entries in $M$) that barely contribute to the $n$-point functions. We will follow this approach in accompanying work when analysing experimental data \cite{ShortPaper}.

\subsection{Reconstructing the poles and residues}\label{sub:recon_poles_res}

When analysing spectra of sampled linear combinations of sinusoidal functions, methods based on integral transforms like the discrete Fourier transform (DFT) seem like a natural choice. In our case, however, we deal with exponentially damped sinusoids with potentially similar frequencies, which results in heavy broadening and overlapping of the corresponding spectral peaks. In this case, the damping factors would have to be determined from the corresponding peaks' width, and, in view of experimental data, we cannot assume too many sampled data points. Hence, the spectral resolution would be rather low. Only for certain cases the peaks in the frequency spectrum are sufficiently separated to directly determine the poles in a feasible way using integral transforms. 

Another class of methods for data fitting that may come to mind is based on non-linear (e.g., least squares) minimisation approaches. Clearly, the number of parameters critically determines the computational effort and the successful applicability of the algorithm. The results, however, can be improved by restricting ourselves to a likely parameter region as a result of a preceding Fourier transform. Taking into account the $\Lambda_d$ and $\Xi_{d,\kappa}$ symmetries and assuming normalised $n$-point functions, the number of real parameters can be reduced to $nd^{2}-2$. Only for unambiguous global minima (which is usually not the case for high damping factors in combination with noise) and for very small bond dimension, we obtained satisfactory results in acceptable time.  Least squares approaches for correlation functions with larger $n$ are at best feasible when using $Q$ and $R$ as parameters, otherwise the number of parameters would become too large. In spite of these drawbacks, a least squares algorithm could be used as an additional refinement step with initial values from other procedures, like the ones discussed below; nevertheless the number of parameters is still limiting. On the other hand, if we can only assume a small number of parameters and expect a considerable amount of noise, the least squares method can be a robust alternative. For example, for bond dimension $d=2$, such non-linear least squares approach can be feasibly and successfully used. 

Alternative minimisation methods, e.g., simulated annealing, did not lead to considerable improvements. However, the scaling of the computational effort with the number of parameters can be significantly mitigated using iterative quadratic maximum likelihood (IQML) methods, but the application to correlation functions with $n>2$ is not straightforward~\cite[Sec.\ 1.2.3]{hua2004high}. 

Realising the challenges of solving a non linear estimation problem, it seems logical to exploit the structure of our particular model of the data to see if there are ways to more efficiently solve the estimation problem. It turns out that for data structures that consist of sums of damped oscillatory terms, it is possible to separate the estimation of poles and residues of the function in two different linear estimation processes. In the following sections, we describe two major approaches one can take to achieve such estimation.

\subsubsection{Prony analysis}\label{sub:Prony-analysis}

This technique is used in digital signal processing and its roots go back to a method that was originally established by R. de Prony in 1795 in the context of fluids~\cite{Prony1795}. The main idea is to first recover the poles independently by determining the roots of a polynomial computed from the signal (the correlation function) and then to insert the poles into a system of linear equations for the coefficients, which is in principle solvable with the usual linear algebra procedures. Prony's method is a special case of \emph{linear prediction}\index{Linear prediction}~\cite{hua2004high} and has many further applications, e.g., as the starting point for nearest-neighbour detection of atoms in optical lattices~\cite{PhysRevLett.102.053001,lee1997riemannian}. The original method, however, is very sensitive to noise, so that for working on experimental data we need to use several modifications, which we will describe below. For further summaries and an introduction of the method, see for instance
Refs.\ \cite{Potts:2010:PEE:1739310.1739358,Lobos,Hildebrand1987}.

Prony's method is usually applied to $\cc$-valued functions, corresponding to 2-point functions, and for our purposes has to be extended to work with higher order $n$-point functions, which can be done in a straightforward way. Therefore, in our description, we will start with the one-dimensional case with signal function 
\begin{equation}
	C^{(2)}(\tau):=\sum_{k=1}^{d^2}\rho_{k}e^{\lambda_{k}\tau}.
\end{equation}
The function is sampled at a finite number of points and is available only for $N+1$ points $\{\tau_{j}\}$, which is $C^{(2)}(\tau_{j}):=C_{j}$, $j=0,\dots,N$. We, thus, obtain a system of linear equations
\begin{align} 
\rho_{1}e^{\lambda_{1}\tau_{0}}+\dots+\rho_{d^2}e^{\lambda_{d^2}\tau_{0}} & =C_{0},  \\
\rho_{1}e^{\lambda_{1}\tau_{1}}+\dots+\rho_{d^2}e^{\lambda_{d^2}\tau_{1}} & =C_{1}, \\  & \vdots\nonumber \\
\rho_{1}e^{\lambda_{1}\tau_{N}}+\dots+\rho_{d^2}e^{\lambda_{d^2}\tau_{N}} & =C_{N}.
\end{align}
Once we have identified all poles $\{\lambda_{k}\}$, we can easily solve this system and are finished with the reconstruction. As we will see, one requirement for Prony's method is to sample the signal at equidistant points $\tau_{j}=j\cdot\Delta \tau$, $j\in\{0,\dots,N\}$, and with $e^{\lambda_{k}\Delta \tau}=:\mu_{k}$ we arrive at
\begin{equation}
\begin{pmatrix}1 & 1 & \cdots & 1\\
\mu_{1} & \mu_{2} & \cdots & \mu_{d^2}\\
\vdots & \vdots &  & \vdots\\
\mu_{1}^{N} & \mu_{2}^{N} & \cdots & \mu_{d^2}^{N}
\end{pmatrix}
\begin{pmatrix}\rho_{1}\\
\rho_{2}\\
\vdots\\
\rho_{d^2}
\end{pmatrix}=
\begin{pmatrix}C_{0}\\
C_{1}\\
\vdots\\
C_{N}
\end{pmatrix},\label{eq:prony: vandermonde}
\end{equation}
where the poles are encoded in the (in general, non-square) Vandermonde matrix \begin{equation}
\mbox{\ensuremath{\mathcal{V}}}:=\left(\mu_{k}^{j-1}\right)_{\substack{j=1,\dots,N+1\\
k=1,\dots,d^2~~~}}.
\end{equation}
We must take care not to choose the sampling interval $\Delta\tau$
too large, since, considering the Nyquist-Shannon sampling theorem \cite{Shannon1949},  the sampling rate should in general be at least twice the highest frequency $\omega_{\text{sup}}$ of the 
signal spectrum ${2\pi}/({\Delta\tau})<2\omega_{\text{sup}}$. 

Vandermonde matrices will often be ill-conditioned---e.g., according to Gautschi~\cite{Gautschi1978}, a lower bound for the norm of the inverse matrix of $\mathcal{V}$ (for $N=d^2$ and $\mathcal{V}$ invertible) is
\begin{equation}\left\Vert \mbox{\ensuremath{\mathcal{V}}}^{-1}\right\Vert _{\infty}>\max_{1\leq l\leq n}\prod_{\substack{m=1\\
m\neq l}}^{d^2}\frac{\max\left(1,\left|\mu_{m}\right|\right)}{\left|\mu_{l}-\mu_{m}\right|},\end{equation}
which will get very large if two poles get close to each other. This fact hints at the intrinsic limitations of this reconstruction method.

To determine the poles, we can regard the set $\{ \mu_{1},\dots,\mu_{p}\} $ as the roots of a polynomial $\mathcal{P}_{d^2}$ with real coefficients and degree $d^2$ in the variable $z$,
\begin{align}
&\mathcal{P}_{d^2}(z)  =\sum_{l=0}^{d^2}a_{l}z^{l}\textsf{,}\label{eq:prony:polynomial}\\
&\mathcal{P}_{d^2}(z=\mu_{k})  =0\nonumber 
\end{align}
for each $k=0,\dots,N$. Note that there are $d^2$ values of $\mu_{k}$ but $d^2+1$ of $a_{l}$.
Such a polynomial naturally exists---it is just the product of the
linear factors $(z-\mu_{k})$,
\begin{equation}	
	\mathcal{P}_{d^2}(z)=\prod_{k=1}^{d^2}(z-\mu_{k}).
	\end{equation}
Our goal is to relate the set of coefficients $\{a_{l}\}$ to the set of function
values $\{C_{j}\}$. Once we have all $a_{l}$, we can compute
the roots of the corresponding polynomial~\eqref{eq:prony:polynomial}
and obtain the poles $\lambda_{k}=\ln(\mu_{k})/\Delta \tau$, $k=1,\dots,d^2$.
To this end, we multiply the first line of Eq.~\eqref{eq:prony: vandermonde}
by $a_{0}$, the second by $a_{1}$ and so on, and perform the sum,
\begin{align}
\sum_{l=0}^{d^2}a_{l}C_{l} & =\sum_{l=0}^{d^2}a_{l}\sum_{k=0}^{d^2}\rho_{k}\mu_{k}^{l} =\sum_{k=0}^{d^2}\rho_{k}\sum_{l=0}^{d^2}a_{l}\mu_{k}^{l}.\label{eq:prony:zero}
\end{align}
Now, by choice of the $a_{l}$, each $\mu_{k}$ is a root of $\mathcal{P}_{d^2}(z)$
for all $k=1,\dots,d^2$ so that each sum over $l$ in Eq.~\eqref{eq:prony:zero}
vanishes. Accordingly, we see that 
\begin{equation}
\sum_{l=0}^{d^2}a_{l}C_{l}=0.\label{eq:prony:zero2}
\end{equation}
Since $\prod_{k=1}^{d^2}\left(z-\mu_{k}\right)=1\cdot z^{d^2}+\dots$, the coefficient
$a_{d^2}$ belonging to the highest power is equal to one. Hence, Eq.~\eqref{eq:prony:zero2}
becomes the recurrence relation
\begin{equation}
\sum_{l=0}^{d^2-1}a_{l}C_{l}=-C_{n}.\label{eq:prony:bn=00003D1}
\end{equation}
In order to compute the $d^2$ coefficients $\{a_{0},\dots,a_{d^2-1}\}$,
we need at least $d^2$ equations. More linear independent equations
are easy to obtain because the argument in Eq.~\eqref{eq:prony:zero} is still valid
if we shift $C_{l}$ to $C_{l+m}$ for any $m\in\mathbb{N}$ with
$d^2+m\leq N$:
\begin{align}
\sum_{l=0}^{d^2}a_{l}C_{l+m} & =\sum_{l=0}^{d^2}a_{l}\sum_{k=0}^{d^2}\rho_{k}\mu_{k}^{l+m}\nonumber \\
 & =\sum_{k=0}^{d^2}\rho_{k}\mu_{k}^{m}\left(\sum_{l=0}^{d^2}a_{l}\mu_{k}^{l}\right)=0.\label{eq:prony:zero3}
\end{align}
For $d^2$ equations the largest index that appears is $2d^2-1$ and our equation
system looks like
\setlength{\arraycolsep}{1pt}
\begin{equation}
\begin{pmatrix}C_{0} & C_{1} & C_{2} & \dots & C_{d^2-1}\\
C_{1} & C_{2} & C_{3} &  & \vdots\\
C_{2} & C_{3} & \ddots &  & \vdots \\
\vdots &  &  & \ddots   & C_{2d^2-3}\\
C_{d^2-1} & \dots & \dots  & C_{2d^2-3} & C_{2d^2-2}
\end{pmatrix}
\begin{pmatrix}a_{0}\\
a_{1}\\
\vdots\\
\\
a_{d^2-1}
\end{pmatrix}
=-\begin{pmatrix}C_{d^2}\\
C_{d^2+1}\\
\vdots\\
\\
C_{2d^2-1}
\end{pmatrix}.
\label{eq:prony:hankel}
\end{equation}
Therefore, for $d^2$ poles we need at least $2d^2$ sampling
points $\{C_{0},\dots ,C_{2d^2-1}\}$. The square matrix on
the left hand side of Eq.~\eqref{eq:prony:hankel} can be written
as $(C_{j+k})_{j,k=0,\dots,d^2-1}$ and has the form of a Hankel
matrix. If it is non-singular, the solution vector
$(a_{0},\dots,a_{d^2-1})^{T}$ is unique and can, together with $a_{d^2}=1$,
directly be replaced in~\eqref{eq:prony:polynomial}, 
which in turn will yield the $d^2$ poles in a unique way. Hence, when
reconstructing a function with $d^2$ poles and residues, we need \emph{precisely}
$2d^2$ sampling points to \emph{exactly} solve the Hankel and the Vandermonde
system, provided that both matrices are not singular. This means that
for small bond dimensions and without noise the necessary resolution
of the signal for a complete reconstruction is very low.  

There are many established criteria for the invertibility~\cite[§ 18]{Iohvidov1982}
and inversion algorithms~\cite{ChandrasekaranToeplitz,Trench1964}
of Hankel or Toeplitz matrices (Eq.~\eqref{eq:prony:hankel} can also be rearranged as a Toeplitz system.). They are known to be potentially ill-conditioned, which
reflects the inverse nature of the problem, e.g., the spectral condition
number of a real positive-definite $N\times N$ Hankel matrix is
bounded from below by $3\cdot2^{N-6}$~\cite{Tyrtyshnikov1994}. In practice, recovering the poles is more stable when oversampling the signal and using a higher pole estimate, i.e., working with a larger (not necessarily square) Hankel matrix and a larger solution vector in Eq.~\eqref{eq:prony:hankel}, and solving the equation system in a least squares sense.  This boils down to applying the Moore-Penrose pseudoinverse to the right hand side of Eq.~\eqref{eq:prony:hankel} to obtain the coefficients of the polynomial, inserting the computed poles into Eq.~\eqref{eq:prony: vandermonde} and discarding the $N+1-p$ surplus poles with the smallest associated residues. %\AS{source, the best relation between cols and rows is ... => Hankel NxN}

Note that instead of solving Eq.~\eqref{eq:prony:hankel}, we can also determine
the kernel of\linebreak $(C_{j+k})_{j,k=0,\dots,d^2-1}$,
whose dimension is larger or equal to one due to Eq.~\eqref{eq:prony:zero2}.
Only in the latter case, which corresponds to the matrix in Eq.~\eqref{eq:prony:hankel}
being non-singular, we get a unique (up to multiplication by a constant)
solution vector $(a_{0},\dots,a_{d^2})^{T}$. The
constant does not pose a problem because any multiple of $(a_{0},\dots,a_{d^2})^{T}$
yields the same roots of the associated polynomial: $\sum_{l=0}^{d^2}\alpha ~a_{l}z^{l}=0$
is equivalent to $\sum_{l=0}^{d^2}a_{l}z^{l}=0$. This method has proven
to be more robust towards noise in some cases \cite{Mackay1987} and can be generalised in an elegant way to higher order correlation functions~\cite{SacchiniProny}.

Unfortunately, in many cases, Prony's method is highly susceptible to noise in the signal. However, it presents a beautiful framework that shows that, in principle, it is possible to reconstruct the poles and residues of a signal. Without noise, both poles and residues can be determined {\it exactly}. In the next section, we describe a better algorithm for solving this type of inverse problems, which is more stable for larger bond dimension and finer sample rates.

\subsubsection{Matrix pencil method}\label{Sec:MPM}

The original matrix pencil method (MPM) was developed by Hua and Sarkar~\cite{HuaSarkarGPOF,HuaSarkarMPM1990} and can be directly applied to our problem. As with the Prony algorithm, the poles are determined first and independently from the residues. Although the MPM is related to Prony~\cite{SarkarPereira}, 
it is considerably less sensitive to noise~\cite[Sec.\ 1.2]{hua2004high}
and can deal with higher sampling rates in a more stable fashion. Once the poles are identified, the
residues are found via a linear equation system in the same way as in Prony's method. Here, we will just describe how to determine the poles. For simplicity, we will begin with the case of reconstructing a 2-point function and generalise to higher order correlation functions in the following section.

A \emph{matrix pencil}\index{Matrix pencil} $\mathfrak{M}$
of degree $n\in\mathbb{N}$ is a polynomial over $\cc$ with
matrix valued coefficients $M_j\in\cc^{d\times d}$,
$\mathfrak{M}\left(\gamma\right)=\sum_{j=0}^{n}M_{j}\gamma^{j}$. As with the Prony algorithm, we start
by forming the Hankel matrix
\begin{equation}
C^{[1]}:=\begin{pmatrix}C_{0} & C_{1} & \dots & C_{P-1}\\
C_{1} & C_{2} & \dots & C_{P}\\
\vdots & \vdots &  & \vdots\\
C_{N-P-1} & C_{N-P} & \dots & C_{N-2}
\end{pmatrix}\in\cc^{\left(N-P\right)\times P},\label{eq:MatrixPencil:Hankel}
\end{equation}
from the experimental data points $\left\{ C_{0},\dots ,C_{N-2}\right\}$
\begin{equation}	
	C_{j}=\sum_{k=1}^{d^{2}}\rho_{k}\textrm{e}^{\lambda_{k}\Delta \tau\cdot j}=\sum_{k=1}^{d^{2}}\rho_{k}\mu_{k}^{j},
\end{equation}
with integers $N,P$, such that $N-P,  P>d^{2}$. Generally, the larger the number of samples $N$,
the better the estimation of poles becomes. The optimal value for $P$ regarding noise sensitivity typically lies between $N/3$ and $N/2$~\cite{HuaSarkarSVDMP}.
In this method, we make use of the fact that $C^{[1]}$ can always be decomposed as
\begin{equation}
	C^{[1]}=\mbox{\ensuremath{\mathcal{V}}}_{1} \mathcal{R} \mathcal{V}_{2}\label{eq:matrpencil:Hankel decomp}
\end{equation}
with Vandermonde matrices 
\begin{equation}
\mbox{\ensuremath{\mathcal{V}}}_{1}=\begin{pmatrix}1 & 1 & \dots & 1\\
\mu_{1} & \mu_{2} &  & \mu_{d^{2}}\\
\vdots & \vdots &  & \vdots\\
\mu_{1}^{N-P-1} & \mu_{2}^{N-P-1} & \dots & \mu_{d^{2}}^{N-P-1}
\end{pmatrix}\in\cc^{\left(N-P\right)\times d^{2}}
\end{equation}
and
\begin{equation}
\mbox{\ensuremath{\mathcal{V}}}_{2}=\begin{pmatrix}1 & \mu_{1} & \dots & \mu_{1}^{P-1}\\
1 & \mu_{2} &  & \mu_{2}^{P-1}\\
\vdots & \vdots &  & \vdots\\
1 & \mu_{d^{2}} & \dots & \mu_{d^{2}}^{P-1}
\end{pmatrix}\in\cc^{d^{2}\times P},
\end{equation}
and the diagonal matrix $\mathcal{R}=\textrm{diag}\left(\rho_{1},\dots,\rho_{d^{2}}\right)$,
as can easily be verified by using Eq.~\eqref{eq:matrpencil:Hankel decomp}. 
In addition to the Hankel matrix $C^{[1]}$, we construct a second Hankel matrix 
\begin{equation}
C^{[2]}=\begin{pmatrix}C_{1} & C_{2} & \dots & C_{P}\\
C_{2} & C_{3} & \dots & C_{P+1}\\
\vdots & \vdots &  & \vdots\\
C_{N-P} & C_{N-P+1} & \dots & C_{N-1}
\end{pmatrix}\in\cc^{\left(N-P\right)\times P}\textsf{,}
\end{equation}
which in turn can be decomposed as 
\begin{equation}
	C^{[2]}=\mbox{\ensuremath{\mathcal{V}}}_{1}
	\mathcal{R}
	\mbox{\ensuremath{\mathcal{V}}}_{0}
	\mathcal{V}_{2}
	\label{eq:matrpencil:Hankel decomp 2}
\end{equation}
with $\mbox{\ensuremath{\mathcal{V}}}_{0}=\textrm{diag}\left(\mu_{1},\dots,\mu_{d^{2}}\right)$,
and consider the linear matrix pencil
\begin{equation}
C^{[2]}-\gamma C^{[1]}=\mbox{\ensuremath{\mathcal{V}}}_{1}\mathcal{R}\left(\mbox{\ensuremath{\mathcal{V}}}_{0}-\gamma\one_{d^{2}}\right)\mathcal{V}_{2}
\end{equation}
with $\gamma\in\cc$. Since all $\mu_{j}$ of $\mbox{\ensuremath{\mathcal{V}}}_{1}$
and $\mbox{\ensuremath{\mathcal{V}}}_{2}$ are distinct for a non-degenerate spectrum of $T$ and $N-L,  L>d^{2}$,
the matrices $\mbox{\ensuremath{\mathcal{V}}}_{1}$ and $\mbox{\ensuremath{\mathcal{V}}}_{2}$ have rank $d^{2}$  and we can see that 
\begin{equation}
\textrm{rank}\left(C^{[1]}\right)=\textrm{rank}\left(C^{[2]}\right)=\textrm{rank}\left(\mbox{\ensuremath{\mathcal{V}}}_{1} \mathcal{R} \mbox{\ensuremath{\mathcal{V}}}_{0} \mathcal{V}_{2}\right)=\textrm{rank}\left(\mathcal{R}\right)=d^{2}\textsf{.}
\end{equation}

Generically, the matrix pencil $C^{[2]}-\gamma C^{[1]}$ will have the same rank, except
for $\gamma=\gamma_{j}\in\left\{ \mu_{1},\dots,\mu_{d^{2}}\right\} $.
In that case, the $j$th row of $\left(\mbox{\ensuremath{\mathcal{V}}}_{0}-\gamma\one_{d^{2}}\right)$
is zero, hence 
\begin{equation}
\textrm{rank}\left(C^{[2]}-\gamma C^{[1]}\right)=d^{2}-1\textsf{,}
\end{equation}
and there exists a non-trivial vector $v$ with
\begin{equation}
\left(C^{[2]}-\gamma C^{[1]}\right)v=0\textsf{.}\label{eq:matrpencil:gen EVP}
\end{equation}
In this form, the complex number $\gamma$ can be regarded as a solution of the
\emph{generalised eigenvalue problem}\index{Generalised eigenvalue problem}
(GEVP) \eqref{eq:matrpencil:gen EVP}. This means that the $d^{2}$
non-zero generalised eigenvalues of Eq.~\eqref{eq:matrpencil:gen EVP}
are exactly the exponentiated poles $\textrm{e}^{\lambda_{1}\Delta t},\dots,\textrm{e}^{\lambda_{d^{2}}\Delta t}$.
Eq.~\eqref{eq:matrpencil:gen EVP} can be solved by a generalised
Schur decomposition
of the matrix pair $\{C^{[2]},C^{[1]}\}$ or by solving
the ordinary eigenvalue problem
\begin{equation}
(C^{[1]})^{+}C^{[2]}v=\gamma v
\end{equation}
with the pseudoinverse\index{Moore-Penrose pseudoinverse} $\left(C^{[1]}\right)^{+}$
of $C^{[1]}$~\cite{HuaSarkarMPM1990}. After having determined the poles this way, they can be inserted into a linear equation system to obtain the according residues, as with Prony's method.

\subsubsection{Technical improvements}

Several improvements can be made to the original MPM approach including features from other reconstruction methods, which led to algorithms like Pro-ESPRIT and TLS ESPRIT~\cite{HuaSarkarSVDMP}, which we mention for the sake of completeness. Modifications based on structured low rank approximations \cite{Cadzow:1991:ERS:137856.137875,LuBovik98improvedmatrix} did not lead to significantly better results. Here, we will focus on the so-called state space matrix pencil method, which shows the highest robustness towards noise of all direct MPM descendants~\cite{HuaSarkarSVDMP,hua2004high} and is the one we prefer to implement. 

%\paragraph{State space matrix pencil method.}
In this context, we continue with Eq.~\eqref{eq:matrpencil:gen EVP}, but instead
of solving it directly, we perform additional noise filtering steps via SVD rank truncations \cite{HuaHuSarkarSVD}. Performing separate SVD truncations like in
the original approach has proven to be less robust than performing
a \emph{joint} SVD on $C^{[1]}$ and $C^{[2]}\in\cc^{\left(N-P\right)\times P}$
by
\begin{equation}
\left(C^{[1]},C^{[2]}\right)=U\Sigma V^{\dagger}=:U\Sigma\left(V^{[1]\dagger},V^{[2]\dagger}\right)\label{eq:matrpencil:jointSVD}
\end{equation}
with a unitary matrix $U\in\mathsf{U}\left(N-P\right)$, $\Sigma\in\cc^{\left(N-P\right)\times2P}$ containing the singular values of the concatenated matrices $(C^{[1]},C^{[2]})\in\cc^{\left(N-P\right)\times2P}$,
and $(V^{[1]\dagger},V^{[2]\dagger})\in\mathsf{U}\left(2P\right)$.
Note that $V^{[1]}$ and $V^{[2]}\in\cc^{P\times2P}$ are not
unitary, in contrast to the matrix $(V^{[1]\dagger},V^{[2]\dagger})$,
and are not directly related to the unitary matrices from the separate
SVDs. We insert Eq.~\eqref{eq:matrpencil:jointSVD} into Eq.~\eqref{eq:matrpencil:gen EVP},
yielding
\begin{equation}
\left(C^{[2]}-\gamma C^{[1]}\right)v=U\Sigma\left(V^{[2]\dagger}-\gamma V^{[1]\dagger}\right)v\textsf{,}
\end{equation}
and see that if $\gamma$ is a generalised eigenvalue of the matrix
pair $\{V^{[2]\dagger},V^{[1]\dagger}\}$, then so it is of $\{C^{[2]},C^{[1]}\}$ . Hence, we can just work with $\{V^{[2]\dagger},V^{[1]\dagger}\}$
(or $\{V^{[2]},V^{[1]}\}$ since the set of poles of our n-point functions
is to be closed under complex conjugation), and can completely forget
about the singular values in $\Sigma$. We now filter the signal given in Eq.\ \eqref{eq:matrpencil:jointSVD}
by keeping the $d^{2}$
largest singular values and the corresponding singular vectors of
$V^{[1]\dagger}$ and $V^{[2]\dagger}$: 

\begin{equation}
U\Sigma\left(V^{[1]\dagger},V^{[2]\dagger}\right)\overset{^{\textrm{trunc}}}{\longmapsto}\left(V_{T}^{[1]\dagger},V_{T}^{[2]\dagger}\right)\textsf{.}
\end{equation}
The GEVP we want to solve now is
\begin{equation}
\left(V_{T}^{[2]}-\gamma'V_{T}^{[1]}\right)v=0\label{eq:matrpencil:GEVP filtered},
\end{equation}
with the filtered eigenvalues $\gamma'\in\cc$. Since
$V_{T}^{[1]},V_{T}^{[2]}\in\cc^{P\times d^{2}}$ and $P\gg d^{2}$,
there is still surplus information we can use to SVD filter Eq.~\eqref{eq:matrpencil:GEVP filtered}
one more time. For higher robustness, we repeat the truncation process, applying it to the concatenated matrix $(V_{T}^{[1]},V_{T}^{[2]})\in\cc^{P\times2d^{2}}$,
\begin{equation}
\left(V_{T}^{[1]},V_{T}^{[2]}\right) =U'\Sigma'\left(V^{\prime[1]\dagger},V^{\prime[2]\dagger}\right)\overset{^{\textrm{trunc}}}{\longmapsto} U'_{T}\Sigma'_{T}\left(V_{T}^{\prime [1]\dagger},V_{T}^{\prime [2]\dagger}\right)
\end{equation}
with $U'\in\mathsf{U}\left(P\right)$, $\Sigma'\in\cc^{P\times2d^{2}}$,
$V'\in\mathsf{U}\left(2d^{2}\right)$, $V'_{T}\in\cc^{d^{2}\times2d^{2}}$
and $V_{T}^{\prime[1]},V_{T}^{\prime [2]}\in\cc^{d^{2}\times d^{2}}$.
Eq.~\eqref{eq:matrpencil:GEVP filtered} then becomes 
\begin{equation}
V_{T}^{[2]}-\gamma'V_{T}^{\prime[1]} =U'\Sigma\left(V^{\prime[2]\dagger}-\gamma'V^{\prime[1]\dagger}\right)
 \mapsto U'_{T}\Sigma'_{T}\left(V_{T}^{\prime[2]\dagger}-\gamma''V_{T}^{\prime[1]\dagger}\right)
\end{equation}
with the doubly SVD filtered eigenvalues $\gamma''\in\cc$.
If there is no noise, then all the $d^{2}$ generalised eigenvalues of the matrix pencil
$\{V_{T}^{\prime[2]},V_{T}^{\prime[1]}\}$ are generalised eigenvalues of $\{V^{\prime[2]},V^{\prime[1]}\}$,
thus generalised eigenvalues of $\{C^{[2]},C^{[1]}\}$ and nothing else than the
exponentiated poles $\textrm{e}^{\lambda_{1}\Delta \tau},\dots,\textrm{e}^{\lambda_{d^{2}}\Delta \tau}$.
With noise, we can assume that the filtered set of eigenvalues $\{\gamma''\}$
provide a better estimate than the unfiltered $\{\gamma\}$~\cite{HuaSarkarSVDMP,HuaHuSarkarSVD}.
Since $V_{T}^{\prime[1]\dagger}$ is invertible by construction, everything
boils down to solving an ordinary eigenvalue problem:
\begin{equation}
\left(V_{T}^{\prime[1]}\right)^{-1}V_{T}^{\prime[2]}v=\gamma''v\textsf{.}
\end{equation}
This concludes the description of the state space matrix pencil method, which is our preferred technique for pole reconstruction.

\subsubsection{Generalisation to higher dimensions}
So far, we have developed the reconstruction techniques for 2-point correlation functions. In this section, we show how to deal with higher order functions and generalise the previous discussion. Additionally, we show how one can improve the signal-to-noise ratio by exploiting redundant information in the higher order correlation functions.

If, for an $n$-point function, we uniformly sample each tensor index with $N$ sampling points, we obtain a $\left(n-1\right)$-dimensional array $\left(C_{l_{1},\dots,l_{n-1}}\right)_{l_{1},\dots,l_{n-1}=0,\dots,N-1}\in\cc^{N^{n-1}}$ with
\begin{equation}
C_{l_{1},\dots,l_{n-1}}  =\sum_{k_{1},\dots,k_{n-1}=1}^{d^{2}}\hspace{-1em}\rho_{k_{1},\dots,k_{n-1}}^{\left(n\right)}\textrm{e}^{\lambda_{k_{1}}l_{1}\Delta\tau}\cdots\textrm{e}^{\lambda_{k_{n-1}}l_{n-1}\Delta\tau}\label{eq:prony n dim gen}
\end{equation}
To extract the poles, we carry forward the approach of Zhu and Hua~\cite[chap.\ 17.11]{yan2002signal}. We fix one index $l_j$ of $C_{l_{1},\dots,l_{n-1}}$ and sum over the other indices 
\begin{equation}
\hat C_{l_j}^{(j)}:=\sum_{\substack{\{l_i\}=0,\\ i\neq j}}^{N-1}C_{l_{1},\dots,l_{n-1}}.
\end{equation}
The summing provides averaging and hence increases noise stability. This procedure is only possible because the poles and the sampling interval are the same for each index of the $n$-point function data array. Inserting the definition for $C_{l_{1},\dots,l_{n-1}}$ and separating $\textrm{e}^{\lambda_{_{k_{j}}}l_{j}\Delta\tau}$ from the summation of $k_j$ yields
\begin{equation}\label{eq:res_ndim}
\hat C_{l_j}^{(j)}=\sum_{k_{j}=1}^{d^{2}} \check C_{k_j}^{(j)} \textrm{e}^{\lambda_{_{k_{j}}}l_{j}\Delta\tau}
\end{equation}
with
\begin{equation}
\check C_{k_j}^{(j)} = \sum_{\substack{\{k_i\}=1,\dots, d^2,\\\{l_i\}=0,\dots,N-1\\i \neq j}}\hspace{-1em}\rho_{k_{1},\dots,k_{n-1}}\textrm{e}^{\lambda_{k_{1}}l_{1}\Delta\tau}\cdots\textrm{e}^{\lambda_{_{k_{j-1}}}l_{j-1}\Delta\tau}\textrm{e}^{\lambda_{_{k_{j+1}}}l_{j+1}\Delta\tau}\cdots\textrm{e}^{\lambda_{k_{n-1}}l_{n-1}\Delta\tau}
\end{equation}
Eq.~\eqref{eq:res_ndim} can be be regarded as the components of a  2-point function with the sought-after poles and $\{\check C_{k_j}^{(j)}\}$, which only depend on $k_j$, as its residues. The concrete values of these effective residues do not matter, since in this step we are only interested in the poles.
We can average further by summing the vectors $(\hat{C}_{l_j}^{(j)})_{l_j=0,\dots,N-1}$, each corresponding to the tensor direction $j$, which leads to the $N$-component vector
\begin{equation}
	\hat{(C_l)}_l:=(\hat{C}_{l}^{(1)})_l+(\hat{C}_{l}^{(2)})_l+\dots+(\hat{C}_{l}^{(n-1)})_l.
\end{equation}
The counting indices $\{l_j\}$ do not depend on $j$, hence we omitted the $j$ for clearness. 

The vector $(\hat{C}_{l})$ still corresponds to a 2-point function with the correct poles and we can now apply the established matrix pencil, Prony or a least squares method to obtain the poles.
Additionally, the averaging results in an effective reduction
of the standard deviation of the (white) noise by a factor of $((n-1)N^{n-1})^{-1}$.
Regarding the residues, we can reshape the array of the poles into a matrix and obtain the residues as the solution vector of the corresponding linear equation system in the least squares sense.

%%%%%%

\subsection{Extracting $M$}
After having determined the poles and residues of the input correlation function---our first reconstruction step as discussed in Sec.\ \ref{sec:steps}---the next step is to identify the matrix $M$. From $M$ together with $D$, the variational parameter matrices $R$ and $Q$ can be determined. %Even when only knowing $M$, it is already possible to generate $n$-point functions for arbitrary $n$. This is especially true if some poles effectively are screened by noise and only an effective $M$ with dimension lower than $d^2$ is obtainable. %\AS{[This is already mentioned in more detail in "reconstruction steps".]}

First, we note that conjugating $M$ with a diagonal matrix whose first entry is equal to one does not change the density-like correlation functions. This observation can be used to require that $M_{1,j}=1$ for $j=2,\dots,d^2$, which is possible if the $M_{1,j}$ are non-zero. For $M_{1,1}$ to be equal to one, we need to normalise the $n$-point function by dividing by 
\begin{equation}
	\langle\Psi_{Q,R}|\hat{\Psi}^{\dagger}\hat{\Psi}|\Psi_{Q,R}\rangle^{n}=M_{1,1}^{n}.
\end{equation}
In particular, we obtain $\rho_{k_1,1,\dots,1}^{(n)}=1\cdot\dots\cdot1\cdot M_{k_{1},1}=\rho_{k_{1}}^{(2)}$. For clearness, in this section we mark the dimensions of the residues with an additional index.
We can compute $M_{i,j}$  for any $i,j=1,\dots,d^{2}$ and $n\geq3$ via 
\begin{equation}
	\frac{\rho_{j,i,1,\dots,1}^{(n)}}{\rho_{j,1,\dots,1}^{(n)}}=\frac{\rho_{j,i}^{(3)}}{\rho_{j}^{(2)}}=\frac{M_{i,j}M_{j,1}}{M_{j,1}}=M_{i,j}.
\end{equation}		
From this equation we can see that we need $n$ to be larger than three, since a 2-point function can at best provide the first column of $M$.

In practice, we may want to reduce noise by averaging over multiple independent prescriptions for $M_{i,j}$, namely 
\begin{equation}
	M_{i,j} =\frac{1}{d^{2\left(n-3\right)}}\sum_{k_{1},\dots , k_{n-3}=1}^{d^{2}}\frac{\rho_{k_{1},\dots,k_{n-3},j,i}^{(n)}}{\rho_{k_{1},\dots,k_{n-3},j,1}^{(n)}}.
\end{equation}	
%From this, any other $n$-point function \eqref{eq:expval} can be generated.
By rearranging the residues, we can express higher order expectation values in terms of lower order:
\begin{equation}
\begin{split}
\rho_{k_{1},\dots,k_{n-1}}^{(n)}=& M_{1,k_{n-1}}M_{k_{n-1},k_{n-2}}M_{k_{n-2},k_{n-3}}\dots M_{k_{1},1}\\=&M_{1,k_{n-1}}M_{k_{n-1},k_{n-2}}\frac{M_{k_{n-2},1}M_{1,k_{n-2}}}{M_{1,k_{n-2}}M_{k_{n-2},1}}M_{k_{n-2},k_{n-3}}\\&\cdots\frac{M_{k_{2},1}M_{1,k_{2}}}{M_{1,k_{2}}M_{k_{2},1}}M_{k_{2},k_{1}}M_{k_{1},1}\\=&\rho_{k_{1},k_{2}}^{(3)}\prod_{r=2}^{n-2}\frac{\rho_{k_{r},k_{r+1}}^{(3)}}{\rho_{k_{r}}^{(2)}}.
\end{split}
\end{equation}
This is the Wick's theorem for matrix product states \cite{Wick_MPS}.
At this point, we can check the validity of the reconstructed $M$, since it necessarily must obey the symmetry $\Xi_{d,\kappa}\overline{M}\Xi_{d,\kappa}=M$ for accordingly ordered spectrum of $T$. 

\subsection{Extracting $R$}
To obtain a complete cMPS description of the system at hand, it is necessary to reconstruct the variational parameter matrices $R$ and $Q$.
We have that, by definition, 
\begin{equation}
	M=X^{-1}\left(\overline{R}\otimes R\right)X 
\end{equation}
and $D=\textrm{diag}\left(\lambda_{j}\right)=X^{-1}TX$ with the change-of-basis matrix $X$  indeterminate.
Because of the gauge invariance of $Q$ and $R$, we can determine them only up to conjugation with an invertible matrix and therefore will not need to determine the concrete form of $X$ at all. In this sense, there are no specific $R$ and $Q$ matrices to be reconstructed. Nevertheless, we continue using the terms $R$ and $Q$, thinking, without loss of generality, of matrices that are in a specific, yet arbitrary, gauge.

Our strategy to recover the variational parameter matrices is to choose $R$ diagonal, which can be done in almost all cases, and determine $Q$ accordingly. Equivalently, one could likewise require $Q$ to be diagonal and determine $R$ accordingly, but here we use the former approach. We first diagonalise $M\mapsto Y^{-1}MY=M_\mathrm{diag}$ with the change-of-basis matrix $Y$. Since $M$, as well as its similar matrix $\overline{R}\otimes R$, has the spectrum $\{\overline{r_i}r_j\}$ with $i,j=1,\dots,d$, where $r_1,\dots,r_d$ are the eigenvalues of $R$, the entries of $M_\mathrm{diag}$ can be reordered with a permutation matrix $O$  such that the resulting matrix has the form of a Kronecker product of two diagonal matrices $R_{\textrm{rec}}$ 
\begin{equation}\label{eq:M_diag}
O^{-1}M_\mathrm{diag}O=\overline{R}_{\textrm{rec}}\otimes R_{\textrm{rec}}.
\end{equation}
Since $R_{\textrm{rec}}$ by construction is similar to $R$, we can write it as $R_{\textrm{rec}}=W^{-1}RW$, where $W$ is the change-of-basis matrix that diagonalises $R$. Diagonalising and reordering $M$ thus yields $R$ in a certain gauge, namely $W^{-1}RW$, and we can identify $R_{\textrm{rec}}$ with a reconstruction of the matrix $R$. 

Note that $XYO$ has a Kronecker product structure as well, which will be important for reconstructing $Q$. Rewriting Eq. \eqref{eq:M_diag}, we have
\begin{equation}
\left(XYO\right)^{-1}\left(\overline{R}\otimes R\right)XYO=O^{-1}Y^{-1}MYO
\end{equation}
which is equal to $\overline{R}_{\textrm{rec}}\otimes R_{\textrm{rec}}$, and, by definition of $R_{\textrm{rec}}$ and using a Kronecker product identity, hence equal to  
\begin{equation}\label{eq:WW}
\left(\overline{W}\otimes W\right)^{-1}\left(\overline{R}\otimes R\right)\left(\overline{W}\otimes W\right).
\end{equation} 
There is a little subtlety in that, in general, numerical diagonalisation algorithms will not provide $Y$ such that $XYO$ is a Kronecker product, but usually such that each eigenvector, a column of $Y$, is normalised, yielding a matrix $Y_N$. This matrix can also be written as $Y_N=YD_Y$ with a diagonal matrix $D_Y$, where $XYD_YO$ in general will not correspond to a Kronecker product. %\AS{This relates to the fact, that Eq. \eqref{eq:M_diag} to \eqref{eq:WW}, being diagonal, are invariant }
This does not affect $R_\textrm{rec}$, since diagonal matrices are invariant under conjugation with other diagonal matrices. 

To determine $O$ and extract $R_{\textrm{rec}}$ from $\overline{R}_{\textrm{rec}}\otimes R_{\textrm{rec}}$, it is important to take into account that multiplying $R$ with an  arbitrary complex phase factor $\textrm{e}^{\textrm{i}\varphi}$ does not change $\overline{R}\otimes R$. In the same way, $\overline{Q}\otimes\one_{d}+\one{}_{d}\otimes Q$ is left invariant when adding $\textrm{i}\chi\cdot\one_{d}$ with $\chi\in\mathbb{R}$ to $Q$. Hence, the transfer matrix remains unchanged as well. Clearly, out of density-like correlation functions, $R$ and $Q$ can only be reconstructed up to these factors since $Q$ and $R$ only appear in these Kronecker product terms. 

By fixing $\textrm{e}^{\textrm{i}\varphi}$, one diagonal entry $r_j$ of $R_{\textrm{rec}}$ can be assumed to be real and $M_\mathrm{diag}$ can be rearranged to a Kronecker product by successively checking if for an entry $M_{\mathrm{diag},l,l}$ the fraction $\left|M_{\mathrm{diag},l,l}/r_{j}\right|^{2}$ yields another (real) entry of $M_\mathrm{diag}$ (or, in practice with noise, is sufficiently close to it), which must be the case for a Kronecker product matrix with spectrum $\{\overline{r_i}r_j\}$. After repeating this procedure for all entries of $M_{\mathrm{diag}}$, all eigenvalues $\{r_j\}$ are determined, in a fixed order that determines the order of $R_\textrm{rec}$ and $O$ as well. Now, it remains to determine $Q$, which will be done in the next section.
%We see that the reordering operations corresponds to the matrix $O$ and we obtain all values of $R_{\textrm{rec}}$ along the way.

\subsection{Extracting $Q$}\label{sec:extrQ}
The second parameter matrix to be reconstructed, $Q$, will in general not be diagonal in the same gauge where $R$ is diagonal. The goal is to find $Q$ in the appropriate gauge. 
First, we take the matrix $D$, which contains the eigenvalues of $T$, subtract the reconstructed matrix $M$, and see that in principle all the information about $Q$ is stored here:
\begin{align}\label{eq:D-M}
D-M & =X^{-1}TX-X^{-1}\cdot\overline{R}\otimes R\cdot X\nonumber\\
 & =X^{-1}\left(\overline{Q}\otimes\one_{d}+\one{}_{d}\otimes Q\right)X.
\end{align}
By conjugating this with the matrix $YO$, which is the same change-of-basis matrix that directly led from $M$ to $\overline{R}_{\textrm{rec}}\otimes R_{\textrm{rec}}$, we obtain
\begin{align}\label{QQrec}
 & \left(XYO\right)^{-1}\left(\overline{Q}\otimes\one_{d}+\one{}_{d}\otimes Q\right)XYO\nonumber\\
 & =\left(\overline{W}\otimes W\right)^{-1}\left(\overline{Q}\otimes\one_{d}+\one{}_{d}\otimes Q\right)\overline{W}\otimes W\nonumber\\
 & =\overline{W^{-1}QW}\otimes\one_{d}+\one{}_{d}\otimes\left(W^{-1}QW\right)
\end{align}
We obtain in this way $Q_{\textrm{rec}}:=W^{-1}QW$ in the gauge corresponding to the gauge of $R_{\textrm{rec}}=W^{-1}RW$ and thus it represents a valid set of parameters that define the state.
To extract $Q_{\textrm{rec}}$ out of Eq.\ \eqref{QQrec}, we can, as in the case of $R_{\textrm{rec}}$, assume one diagonal entry $q_{j,j}$ of $Q_{\textrm{rec}}$ to be real, which corresponds to subtracting $\im \mathfrak{Im}(q_{j,j})\cdot\one_d$ from $Q$. In this way, we can read each $q_{j,j}$ from the corresponding diagonal entry $\overline{q}_{j,j}+q_{j,j}=2q_{j,j}$ in Eq.\ \eqref{QQrec} and subsequently the remaining diagonal entries. Because of the structure of Eq.\ \eqref{QQrec} as a Kronecker sum, the off-diagonal entries can be read off without further preparation. 

The fact that $Y$ is only determined up to multiplication with a diagonal matrix $D_Y$, as mentioned in the previous section, does not pose an obstacle for the reconstruction of $Q_\textrm{rec}$: Its gauge needs to be fixed only up to conjugation with a diagonal matrix if $R_\textrm{rec}$ is in a diagonal gauge. Furthermore, it does not matter that also the matrix $M$ is only determined up to conjugation with a diagonal matrix $D_{M}$, which we used to require that $M_{1,j}=1$ for $j\geq 2$.
Using $D_M^{-1}MD_M$ instead of $M$ in Eq.~\eqref{eq:D-M} and $X^{-1}TX$ being diagonal, we have 
\begin{align}
X^{-1}TX-(XD_{M})^{-1}(\overline{R}\otimes R)XD_{M}=(XD_{M})^{-1}(T-\overline{R}\otimes R)XD_{M},
\end{align}
which is equal to
$\tilde{X}^{-1}\left(\overline{Q}\otimes\one{}_{d}+\one_{d}\otimes Q\right)\tilde{X}$ with $\tilde{X}=XD_M$. The particular structure of $X$ or $\tilde{X}$ is not needed in
the algorithm. 

On the other hand, if we normalise the $n$-point function and hence $M$ by
multiplying it by a constant, we have to be careful since  $D-cM$, for some $c\in\mathbb{R}$, will in general not result
in a matrix similar to $\overline{Q}\otimes\one_{d}+\one{}_{d}\otimes Q$.
Accordingly, we have to \emph{renormalise} $M\mapsto\hat{M}_{1,1}\cdot M$.
The number $\hat{M}_{1,1}$ can be read off the residue $\hat{\rho}_{1,\dots,1}^{\left(n\right)}=(\hat{M}_{1,1})^{n}$
of the $n$-point function before normalising it. 

Note that computing eigenvectors, which the matrix $X$ consists of, can be a very unstable (in extreme cases even discontinuous) procedure, especially for higher bond dimensions, when eigenvalues can cluster \cite[cor. 7.2.6]{golubvanloan1996matrix}. Hence the procedure of determining $Q$ is highly susceptible to noise. To improve noise stability, we can average $Y$ by using the symmetry property $\Xi_{d,\kappa} \overline{Y}\Lambda_d=Y$, which follows from the symmetries of $M$ and  $\overline{R}_{\textrm{rec}}\otimes R_{\textrm{rec}}$, and use $(Y+\Xi_{d,\kappa} \overline{Y}\Lambda_d)/2$ instead. 
%Beyond that, since each $Q_{\textrm{rec}}$ entry in principle appears in $d$ copies in Eq.\ \eqref{QQrec}, these entries can be averaged.

This concludes the reconstruction of the variational parameter matrices $Q$ and $R$, which is the last step in our reconstruction procedure, Sec.\ \ref{sec:steps}. Additionally, it is now possible to construct the Hamiltonian of the auxiliary system $K$ as in Eq.~\eqref{eq:KQ} et sqq.\ and relate the cMPS to a Lindblad master equation. The fact that we can reconstruct $Q$ only up to an additive term $\mathrm{i}\chi\cdot \id1$ results in $K$ being indeterminate up to an additive term $\chi\cdot \id1$. This is reasonable since only the differences in the spectrum of the Hamiltonian are physically relevant and these are not affected by a global shift by $\chi$.

%%%%%%%%%%%%%%%%%%%%%%%%%%%%%%%%%%%%%%%%%%%%%%%%%%%%%%%%%%%%%%%%%%%%%%%%%%%%%%
%%%%%%%%%%%%%%%%%%%%%%%%%%%%%%%%%%%%%%%%%%%%%%%%%%%%%%%%%%%%%%%%%%%%%%%%%%%%%%
%%%%%%%%%%%%%%%%%%%%%%%%%%%%%%%%%%%%%%%%%%%%%%%%%%%%%%%%%%%%%%%%%%%%%%%%%%%%%%

\section{Applicability and limitations}\label{sec:limitations}

The proposed tomography method relies on assumptions. It is hence important to know its limitations and how to check the applicability of the method to given data. The basic assumption is that the correlations in the data are---at least approximately---of the type found in cMPS spatially, or equivalently of the type found in finite dimensional quantum systems whose dynamics are given by a Lindblad equation temporally. It is hence natural to assume that our method is applicable to settings similar to the ground states of gapped local Hamiltonians and for fields which originate from an interaction with finite level systems---think, e.g., of a light beam emitted by an atom trap. In this section, we aim to give a description of ways to gain confidence and check the consistency of the estimates obtained by our reconstruction methods for quantum fields.

Since it is our goal to produce usable estimation tools for experimental applications, it is very important to have a clear understanding of how to determine whether or not a particular reconstruction was successful or even if the cMPS ansatz is applicable to a particular situation. In this context, we can recognize two different scenarios that can occur: 1.~the idealised case, where the data actually comes from a cMPS, and 2.~a realistic case, in which the data comes from a physical system (not a cMPS, but possibly well approximated by one) and is in general noisy. In the following, we will discuss both in more detail.

In the ideal case, data will be produced by a generic cMPS of unknown bond dimension $d$. From the 2-point correlation function, following the reconstruction methods discussed in Sec. \ref{Sec:MPM}, we can extract an estimation of $d$ by computing the rank of the (sufficiently sized) ansatz Hankel matrix in Eq.~(\ref{eq:MatrixPencil:Hankel}). Even if noise is present in the signal, an estimation of the bond dimension can be obtained, because noise-induced singular values are small. Since some of the elements of matrix $M$ can be zero, some of the residues $\rho$ corresponding to poles $\lambda$ can also be zero, thereby hiding those poles. Correlators with different $n$, on the other hand, can reveal these poles at some point, but not necessarily so. Having found all the poles there are, also implying access to the whole matrix $M$, is indicated by an agreement of the poles of all available $n$-point functions. One should keep in mind, though, that one will never be able to verify this, even in the idealised case, with a finite amount of data, as it possible to construct a state which agrees with a given cMPS on e.g., a finite number of $n$-point functions but differs elsewhere. However, a non-increase of the set of poles over a wide range of $n$-point functions is sufficient to build confidence in the correctness of the reconstruction. It is a satisfactory feature of our method that we can quantify the confidence of the reconstruction in this way. 

In contrast, a priori information about the number of expected poles and a guarantee that the number and numerical values of residues and poles will be consistent for all $n$-point functions is not available in most real-world tomographic settings. In fact, when data comes from an experiment, we expect a description in terms of cMPS to be possible only in an approximate sense. A similar situation is known for discrete MPS in a lattice setting, where an exact description of a state can be found only if its Schmidt rank is finite. However, many states whose Schmidt numbers form a fast decaying sequence allow for an efficient description with discrete MPS. Even if the physical system is well approximated by a cMPS in this sense, in general we expect to have an infinite number of poles to recover. However, only a small number of them will be associated to residues that are big enough to contribute to the correlation functions. The number of relevant residues and poles can be identified by looking for singular values of Hankel matrix Eq.~\eqref{eq:MatrixPencil:Hankel} greater than an appropriate threshold. The tomographer, hence, has to formulate a hypothesis about the relevance of the observed poles and try to gain confidence in his/her assumption. The desired situation to observe in practice is that the recovered poles do not change too much (i.e., they are within some threshold, e.g., previously determined by the noise level) independently of the correlation function used to extract them. 

In summary, if the set of poles has to be extended time and again over a wide range of correlation functions, the assumption that the state can be described by a cMPS is clearly wrong. In particular, such a situation would tell us that the cMPS ansatz is not a good model for the particular system and data set. Along the lines of the discussion above, in practice, what we propose to check and gain confidence of the applicability of our methods is the following. Use lower order correlation functions to extract a cMPS description of the system, use the reconstructed cMPS to predict higher order functions and compare them to available measured ones. This way, we can check the consistency of the reconstruction procedure and the validity of the cMPS ansatz for the field state under investigation.

%%%%%%%%%%%%%%%%%%%%%%%%%%%%%%%%%%%%%%%%%%%%%%%%%%%%%%%%%%%%%%%%%%%%%%%%%%%%%%
%%%%%%%%%%%%%%%%%%%%%%%%%%%%%%%%%%%%%%%%%%%%%%%%%%%%%%%%%%%%%%%%%%%%%%%%%%%%%%
%%%%%%%%%%%%%%%%%%%%%%%%%%%%%%%%%%%%%%%%%%%%%%%%%%%%%%%%%%%%%%%%%%%%%%%%%%%%%%

\section{Applications}\label{sec:applications}
 
In this section, we show how the formalism developed so far can be applied to real world scenarios. We demonstrate the applicability in two basic settings. First, we generate correlation functions similar to data obtainable in current experimental settings. For this, we use simulated data to study the performance of the reconstruction method in situations in which noise is present. Second, we analyse the applicability of our techniques to the Lieb-Liniger model, which is a well-known and well-investigated model in one-dimensional non-relativistic field theory.

%%%%%%%%%%%%%%%%%%%%%%%%%%%%%%%%%%%%%%%%%%%%%%%%%%%%%%%%%%%%%%%%%%%%%%%%%%%%%%
%%%%%%%%%%%%%%%%%%%%%%%%%%%%%%%%%%%%%%%%%%%%%%%%%%%%%%%%%%%%%%%%%%%%%%%%%%%%%%

\subsection{Simulations and error analysis}

Before typical noise models can be taken into consideration, we ask what kind of problems we are most likely to encounter. As we have seen, given an arbitrary cMPS $n$-point function with non-degenerate spectrum, its poles and residues can be obtained by matrix pencil or Prony's methods, provided there is sufficient accuracy. We keep in mind that formally it is required that $T$ has a non-degenerate spectrum, which is, however, the case for almost all randomised $T$. Also, it is possible that $M$ contains elements of value zero, which is, likewise, not to be expected. On the other hand, there are other more practical obstacles related to concrete implementation features of the numerical algorithms discussed above. 

%%%%%%%%%%%%%%%%%%%%%%%%%%%%%%%%%%%%%%%%%%%%%%%%%%%%%%%%%%%%%%%%%%%%%%%%%%%%%%

\subsubsection{Typical problems to be expected}

The identification of the poles when determining the matrices $M$ and $D$ is the most critical part of our procedure. More concretely, we face the problem of resolving maxima of the Laplace transform of the correlations in the complex plane. We do not do this directly, but the challenges remain the same. 

The problem is to discern poles that lie close to each other and to identify poles that have comparatively small residues. Moreover, we might face large damping factors, which results in a broadening of the peaks in the Fourier spectrum. The required accuracy for the correct identification of poles and residues hence critically depends on the position of the poles $\{ \lambda_{j}\}$ in the complex plane and the ratio between damping factor $\mathfrak{Re}(\lambda_{j})$ and frequency $\mathfrak{Im}(\lambda_{j})$. Not surprisingly, all these issues are aggravated for higher bond dimensions; the $n$-point functions consist of a larger number of oscillatory components, typically in the vicinity of other poles. Moreover, the reconstruction of the residues will also be affected if the poles are close to each other. This happens because the corresponding linear Vandermonde system of equations becomes more ill-conditioned.

When reconstructing $Q$ from the matrix $M$, we face another type of typical problem. Determining $R$ does not lead to significant additional numerical problems since it mainly involves an ordinary diagonalisation procedure, whereas for reconstructing $Q$,  we need the eigenvectors of $M$, which are very susceptible to perturbations of the matrix.

In the following, we want to test the robustness of our method by analysing typical noise cases independently. First, as a preparatory step, we generate typical cMPS. Second, we examine how the reconstruction of the poles is affected by adding noise to the input correlation functions. Third, we survey the reconstructability of $R$ and $Q$ when the input for this reconstruction step, the matrix $M$, is perturbed. Fourth, we study the influence of the presence of additional fields.

%%%%%%%%%%%%%%%%%%%%%%%%%%%%%%%%%%%%%%%%%%%%%%%%%%%%%%%%%%%%%%%%%%%%%%%%%%%%%%

\subsubsection{Generating typical cMPS}

In this section, we give a recipe to generate correlation functions with structural features on a desired length scale, based on a randomisation-ansatz for the $Q$ and $R$ matrices. This is in principle a non-trivial task, as the length scales and damping of the fluctuations are directly derived from the spectrum of $T$, which depends non-linearly on the entries of $Q$ and $R$. 

We make the ansatz of generating $Q$ and $R$ as complex Gaussian random matrices with mean $\mu$ and standard deviation $\sigma$---i.e., real and imaginary part of the entries are independently and identically normally distributed according to $\mathcal{N}(\mu,\sigma)$---and renormalise $Q$ such that all eigenvalues of $T$ have real part $\leq 0$. This results in a roughly uniform distribution of the eigenvalues of $T$ within a disc left of the imaginary axis, which is not entirely unexpected when considering Girko's circular law~\cite{ANU:298726} and the Kronecker product structure of $T$. The damping factors of the poles are of the same magnitude as their frequencies or larger, which is not the case if oscillations are actually to be observed and moreover aggravates the identification of such poles and increases the accuracy requirements.

\begin{figure}[t]
\includegraphics[height=.42\columnwidth]{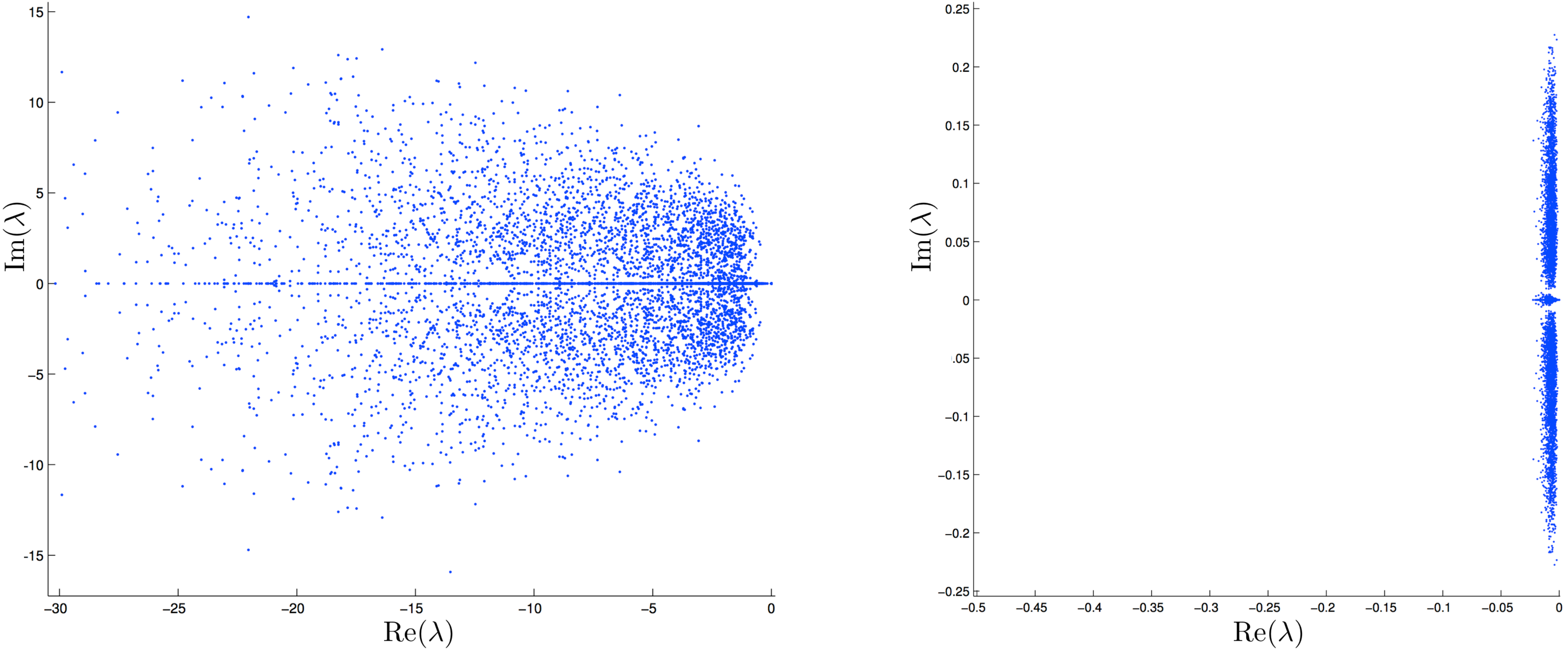}
\caption[Distribution of the poles of 400 cMPS transfer matrices in the complex plane.]{\label{fig:pole_distr}Distribution of the poles of the transfer matrices in the complex plane for 400 cMPS samples with bond dimension $d=4$. The real and imaginary
part of the entries of $K$ and $R$ are i.i.d. with $\mathcal{N}(0,1)$ (a) and $\mathcal{N}(0,0.01)$ (b). In (b), most damping factors corresponding to the real parts of the poles are much smaller than the respective imaginary parts, which correspond to the frequencies components of the correlation functions. This will lead to significantly better reconstructability properties of the cMPS.} 
\end{figure}

In a more refined ansatz, we hence consider sampling $K$ and $R$ instead, from the same distribution, which leads to a drastically higher concentration of poles close to the imaginary axis, when scaling both matrices with a small number $\eta$, see Fig.~\ref{fig:pole_distr}, where we show a comparison of distributions of the poles in the complex plane between the na\"ive and the refined method of randomly sampled cMPS. This scaling of the matrices does not constitute a gauge of the cMPS but rather a transformation to another cMPS, cf.~\cite{PhysRevLett.104.190405}. Matrix $Q$ is mapped to $\frac{1}{2}\eta^{2}R^{\dagger}R-\mathrm{i}\eta K$, see Eq.~\eqref{eq:KQ}, such that for small $\eta$ the eigenvalues of $Q$ will typically feature much larger imaginary part than real part, since the spectrum of $K$ is real and the $R^{\dagger}R$ term adds to $Q$ in second order in $\eta$. This carries over to the construction of $T$ where $\overline{R}\otimes R$ also appears in second order in $\eta$ as opposed to $\overline{Q}\otimes\one+\one\otimes Q$, which are first order. Overall, for small $\eta$ most damping factors become smaller than the frequencies by several orders of magnitude, a property expected to hold if oscillations are observed. Moreover, a distinct peak structure in the Fourier transform emerges, and the poles and residues of $T$ are sufficiently separated and can be determined even with moderate amounts of noise present.

%%%%%%%%%%%%%%%%%%%%%%%%%%%%%%%%%%%%%%%%%%%%%%%%%%%%%%%%%%%%%%%%%%%%%%%%%%%%%%

\subsubsection{Effects of noisy correlation functions}

Typical experimentally measured signals have inaccurate read-out of the signal. We model such noisy situations as Gaussian noise, and study the effect on the reconstruction procedure by adding noise to correlation functions originating from a cMPS. 

In particular, we apply the matrix pencil method to the noisy amputated 2-point function
\begin{equation}
\begin{split}
\hat{C}^{(2)}(\tau_{k})+w(\tau_{k})=\langle \psd(\tau_{k})\psd(0)\ps(0)\ps(\tau_{k})\rangle -\langle \psd(0)\ps(0)\rangle ^{2}+w(\tau_{k}),
\end{split}
\end{equation}
evaluated at $200$ points $\tau_k$, for cMPS with elements  of $R,  K$ sampled from $\mathcal N(0,0.01)$. The white noise function $w$ is sampled from $\mathcal N(0,\textrm{mean}(|\hat{C}^{(2)}|)/\mathrm{SNR})$ where SNR is the signal-to-noise ratio. 

\begin{figure}[t]
\includegraphics[height=.82\columnwidth]{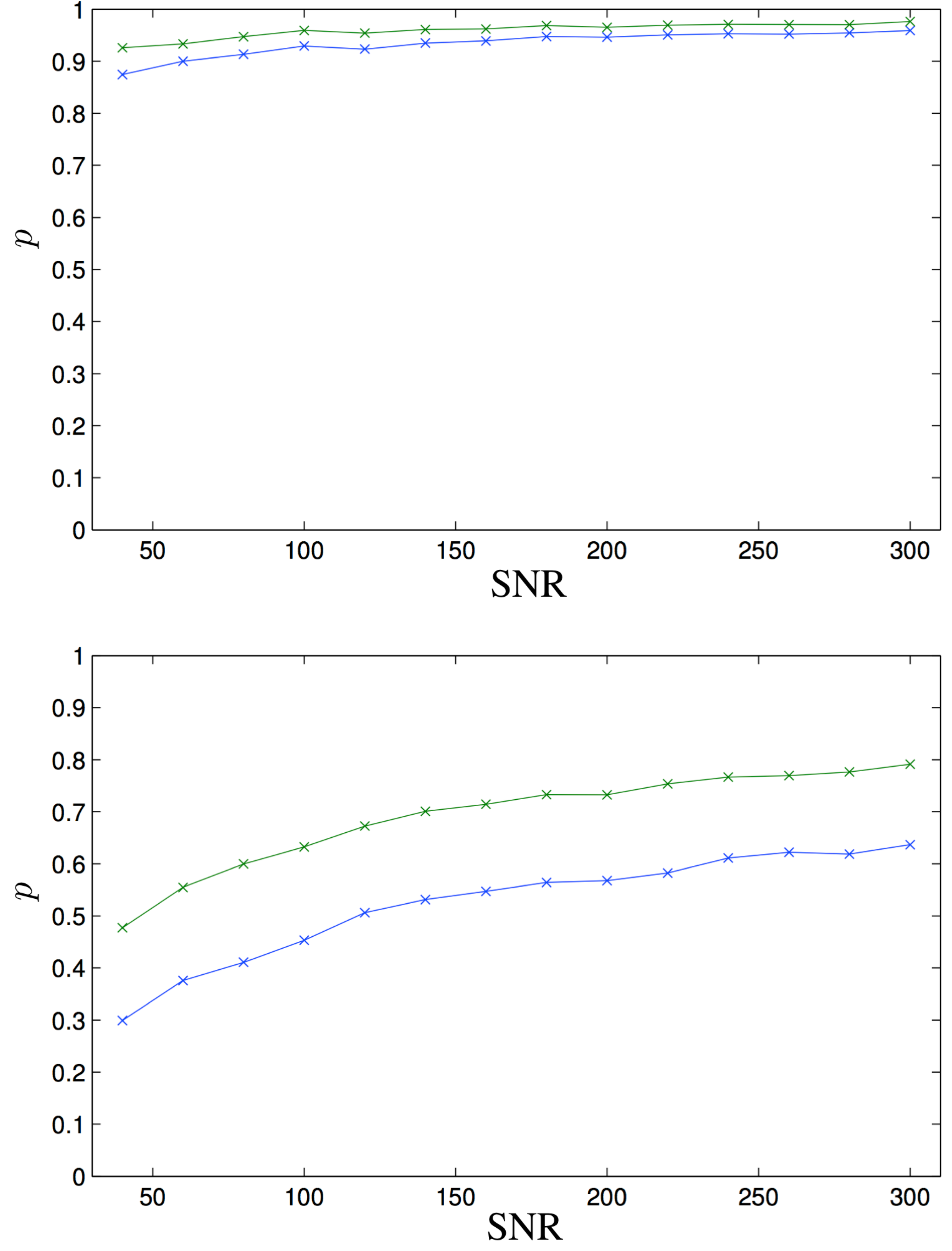}
\caption{\label{fig:recon rate d3}Application of the matrix pencil method
to the signal vector with components $\hat{C}^{(2)} (\tau_{k})+w(\tau_{k}))$ for $d=2$ (above) and $d=3$ (below). $p$ is the percentage of pole sets with $\max_{j=2,\dots,d^{2}} | ({\lambda_{j}- \tilde{\lambda}_{j}})/{\lambda_{j}}|<0.1$ (blue) and $({d^{2}-1})^{-1}\sum_{j=2}^{d^{2}} | ({\lambda_{j}- \tilde{\lambda}_{j}} )/{\lambda_{j}}|<0.1$ (green) as a function of the signal-to-noise ratio, each point summarizing $5000$ runs, where $\{\lambda_{j}\}$ are the original poles, and $\{\tilde{\lambda}_{j}\}$ the pole estimates.}
\end{figure}

In Fig.~\ref{fig:recon rate d3}, $p$ is the percentage of pole sets with $\mathrm{mean}_{j=2,\dots,d^2} |(\lambda_{j}-\tilde{\lambda}_{j})/\lambda_{j}|<0.1$ as a function of the signal-to-noise ratio, where $\{ \lambda_{j}\} $ are the original poles, and $\{ \tilde{\lambda}_{j}\} $ the pole estimates. Each point is computed for $5000$ runs of our numerical experiment to gather enough statistics. What we observe is that for bond dimension $d=2$, our reconstruction procedure is robust to reasonable amounts of noise. However, for bond dimension $d=3$, we see that the robustness is much smaller, which hints to the practical limitations of our reconstruction procedure. The results can, for example, be improved by increasing the sampling rates, however this can be difficult to achieve in experiments. 

Note that in both cases shown in Fig.~\ref{fig:recon rate d3} our procedure behaves as expected from a proper estimator as a function of the SNR: the lesser the noise, the better the reconstruction. In fact, for zero noise, we can in general expect 100\% reconstructability, independent of the bond dimension.
As already mentioned, for higher order correlation functions, $n>2$, the reconstructability of the poles does not necessarily deteriorate---independent of the bond dimension $d$. In fact, since one can average over all projections that fix all but one $\tau$, a significant part of the noise is effectively averaged out.

%%%%%%%%%%%%%%%%%%%%%%%%%%%%%%%%%%%%%%%%%%%%%%%%%%%%%%%%%%%%%%%%%%%%%%%%%%%%%%

\subsubsection{Reconstructability of $Q$ and $R$ when perturbing $M$}

In this section, we look at the next step in the reconstruction process: recovering the cMPS parametrisation matrices $Q$ and $R$ from an imperfectly recovered matrix $M$. We do so by simulating $M$ and perturbing it directly, rather than using a reconstructed $M$ matrix from noisy correlation functions. We do it this way to have control over the size of the perturbation and thus to separate these two different stages of the reconstructed problem and investigate their effect separately. 

For this purpose, we prepare matrices $R$ and $Q$ with entries sampled from $\mathcal N \left(0,1\right)$, then calculate $T$ and $M$, and perturb $M$ with an error matrix $\Delta$. The perturbation has to be carefully designed in order to retain the symmetry $M=\Xi_{d,\kappa} \overline{M} \Xi_{d,\kappa}$ of the unperturbed matrix $M$. This is related to the fact that for any valid reconstruction of a density-like correlation function the residues together with the entries of the matrix $M$ necessarily are either real or appear in pairs of complex conjugates, see Sec.\ \ref{sec:add_sym}. Perturbing with the matrix
\begin{equation}
\Delta:=\frac{1}{2}\left(\Delta_{0}+\Xi_{d,\kappa} \overline{\Delta}_{0} \Xi_{d,\kappa}\right) 
\end{equation}
with real and imaginary parts of the entries of $\Delta_{0}$ sampled from $\mathcal N(0,2^{-1/2} \mathrm{mean} (|M|))$ ensures the required symmetry since $\Delta=\Xi_{d,\kappa} \overline{\Delta} \Xi_{d,\kappa}$. Furthermore, since the first row of $M$ is set to one due to normalisation and this should not be changed for perturbed input, the first row of $\Delta$ is set to zero. 
 
\begin{figure}[t]
\begin{centering}
\includegraphics[height=.43\columnwidth]{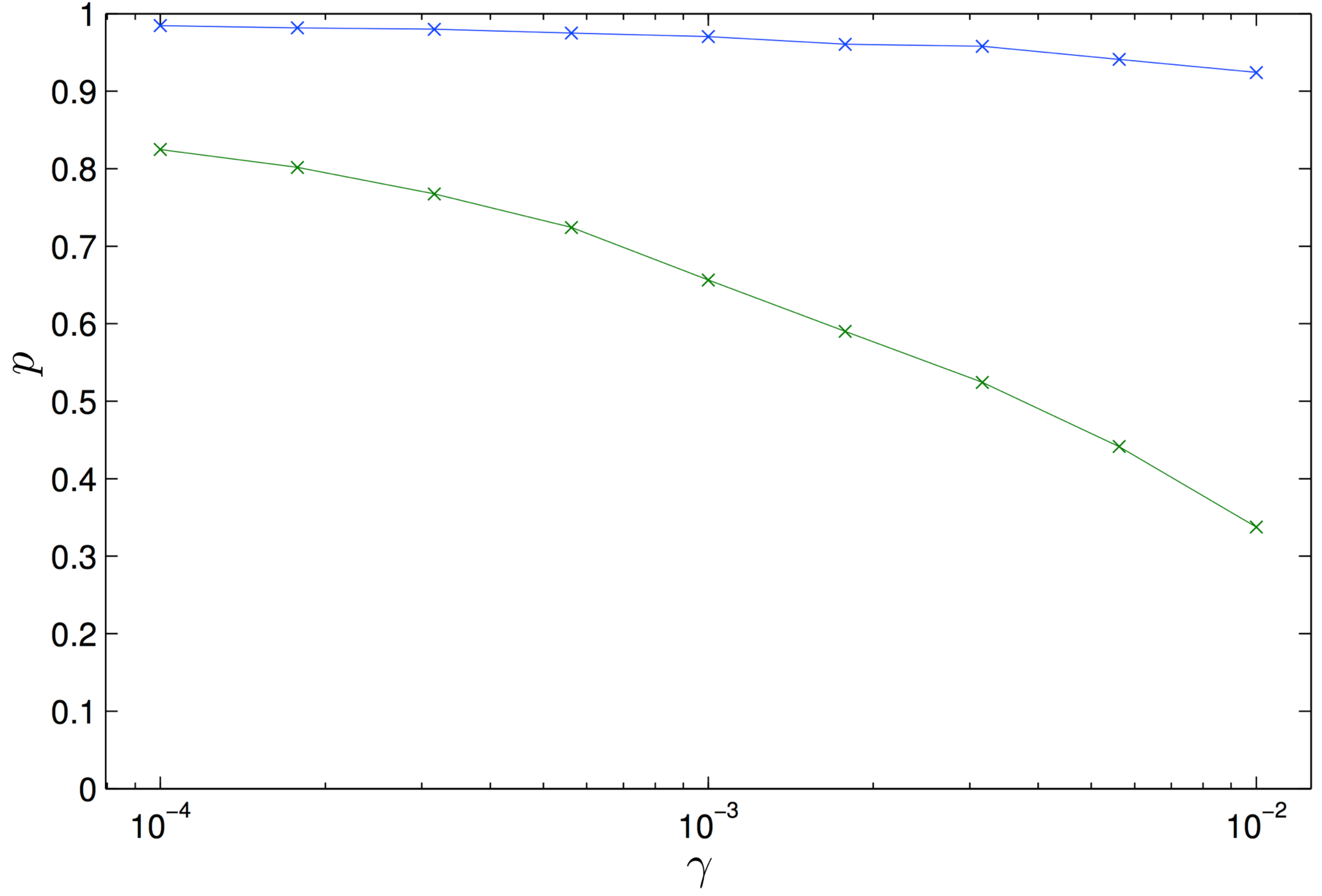}
\end{centering}
\caption[Reconstructability of $Q$ depending on the perturbation of $M$]{\label{fig:applic:recon Q ratio}Reconstructability of $Q$ depending
on the perturbation of $M$: Ratio $p$ out of 5000 samples per point
with $({d^{2}-1})^{-1} \sum_{j=2}^{d^{2}} |({\lambda_{j}-\tilde{\lambda}_{j}})/{\lambda_{j}}|<0.1$
as a function of $\epsilon$ with $\{ \tilde{\lambda}_{j}\} =\sigma(\tilde{T})$
for $d=2$ (blue) and $d=3$ (green). $\tilde{Q}$ and $\tilde{R}$
depend on $\tilde{M}=M+\epsilon\Delta$. As $\epsilon\rightarrow0$ we
have that $p\rightarrow1$.}
\end{figure}

From the reconstructed matrices $\tilde{Q}$ and $\tilde{R}$ from $\tilde{M}=M+\epsilon\Delta$ with scaling parameter $\epsilon\in\mathbb{R}^{+}$ we build the transfer matrix $\tilde{T}$ and compare its spectrum with the spectrum of the original $T$. The ratio of samples with mean deviation $\sigma(\tilde{T})$ to $\sigma(T)$ not larger than 10\% as a function of $\epsilon$ is depicted in Fig.~\ref{fig:applic:recon Q ratio} for bond dimensions $d=2$ (blue) and $d=3$ (green). As the error $\epsilon$ grows, the ratio of successfully reconstructed $Q$ and $R$ matrices drops for both bond dimensions. However, the $d=2$ case is clearly more robust to perturbations. Additionally, we want to point out that any potential deviation of the spectra of $T$ and $\tilde{T}$ is almost certainly due to the reconstruction of $Q$.

%%%%%%%%%%%%%%%%%%%%%%%%%%%%%%%%%%%%%%%%%%%%%%%%%%%%%%%%%%%%%%%%%%%%%%%%%%%%%%

\subsubsection{Effects of additional interactions}

As discussed earlier in Sec.\ \ref{sec:Lindblad}, typical correlations under consideration can be seen as originating from processes where a field state is generated by an interaction with a finite dimensional system, and can be described by a Lindblad equation. In the ideal case, where the finite dimensional system interacts only with the field we measure, we obtain correlations which are perfectly described by a cMPS, or equivalently by a Lindblad equation with one Lindblad operator. In the case where the finite dimensional system interacts with other systems or fields, which we might not even know of, the Lindblad equation is altered and supplemented by more Lindblad operators, which correspond to the other systems or fields. In this case, the transfer matrix takes the form \cite{Osborne2010}
\begin{equation}
T=\mathrm{i}K\otimes\id1-\id1\otimes\mathrm{i}K+\sum_{j}\mathcal{R}_{j}
\end{equation}
where 
\begin{equation}
	\mathcal{R}_{j}=\frac{1}{2}(2\overline{R}_{j}\otimes R_{j}-\overline{R_{j}^{\dagger}R_{j}}\otimes\id1-\id1\otimes R_{j}^{\dagger}R_{j})
\end{equation}
and the additional fields are represented by the terms with $j\geq 2$. Each of the two latter summands in $\mathcal{R}_j$ are connected to $Q$ via Eq.~\eqref{eq:KQ}. The matrix $M$ remains $\overline{R}_1 \otimes R_1$, because it comes from measuring the field corresponding to it, but now in the diagonal basis of a different $T$ than the one for a single field.

\begin{figure}[t]
\begin{centering}
\includegraphics[height=.44\columnwidth]{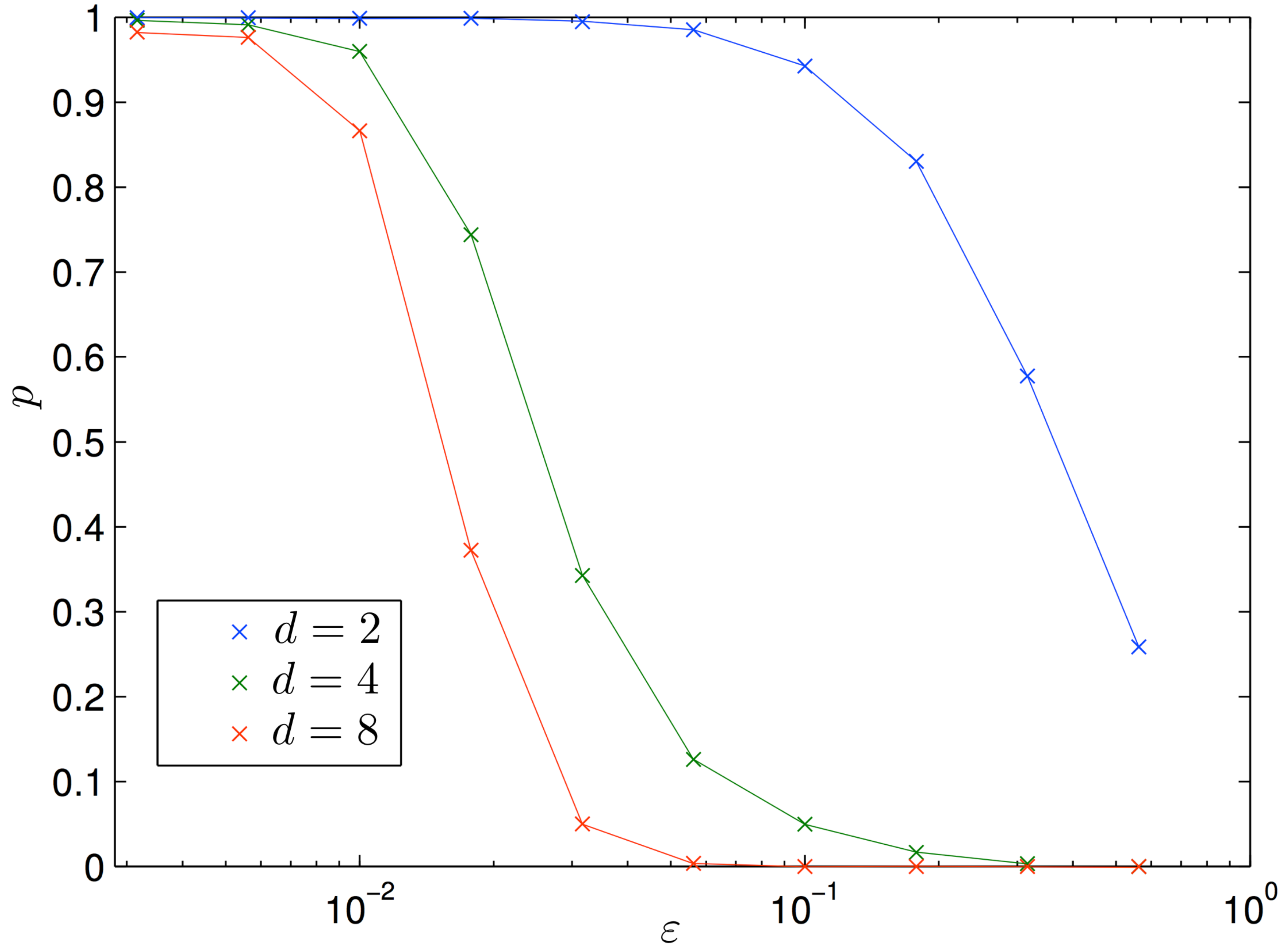}
\end{centering}
\caption{\label{fig:addR}
Reconstruction rate $p$ depending on the size of an additional field and the bond dimension $d$ from $5000$ cMPS samples per point.}
\end{figure}

In order to analyse the sensitivity of reconstructing the variational parameter matrices, we consider one additional perturbation field. More additional fields within the same order of magnitude yield very similar outcomes. This results in $T=\mathrm{i}K\otimes\id1-\id1\otimes\mathrm{i}K+\mathcal{R}_{1}+\epsilon \mathcal{R}_{2}$. In this section, we study how well the spectrum of $K$ can be matched depending on the scaling parameter $\epsilon\in\mathbb{R}^+$. Analogous to the last section, we prepare cMPS by randomly generating $K$, $R_1$, and $R_2$ with elements whose real and imaginary parts are sampled from $\mathcal N(0,1)$. We then generate $M$ matrices and from this reconstruct $R_{1,\mathrm{rec}}$ and an effective $Q_\mathrm{rec}$, assuming only a single field. From $R_{1,\mathrm{rec}}$ and $Q_\mathrm{rec}$ we compute $K_\mathrm{rec}$ and compare the differences of its eigenvalues, $\Delta\tilde{\kappa}_j=\tilde{\kappa}_{j+1}-\tilde{\kappa}_j$, with the differences of the eigenvalues $\kappa_j$ of the actual $K$. Only the differences are reconstructable, see Sec.~\ref{sec:extrQ}. The reconstruction of $K$ is said to be successful if 
\begin{equation}
	\max_{j=1,\dots,d-1}\left|\frac{\Delta\tilde{\kappa}_j-\Delta\kappa_j}{\Delta\kappa_j} \right|<10\%. 
\end{equation}
The reconstruction rate, depending on $\epsilon$ and the bond dimension, is shown in Fig.~\ref{fig:addR}. For $\epsilon\rightarrow 0$ (single field case) all cMPS can be reconstructed. As the size of the additional field approaches the size of the main field, the reconstruction rate drops to zero. The smaller the bond dimension, the more perturbation by additional fields can be tolerated. We conclude that for sufficiently small additional fields, a successful reconstruction is in principle still feasible. Moreover, for $d=2$, the most robust case, this is true even if the additional fields are merely one order of magnitude smaller than the main field.

%%%%%%%%%%%%%%%%%%%%%%%%%%%%%%%%%%%%%%%%%%%%%%%%%%%%%%%%%%%%%%%%%%%%%%%%%%%%%%
%%%%%%%%%%%%%%%%%%%%%%%%%%%%%%%%%%%%%%%%%%%%%%%%%%%%%%%%%%%%%%%%%%%%%%%%%%%%%%

\subsection{The Lieb-Liniger model}

In this section, we analyse the applicability of the results discussed above to the Lieb-Liniger model \cite{1963PhRv..130.1605L}. The model describes the dynamics of a one-dimensional system of bosons interacting via a delta-potential. In second quantisation, the Hamiltonian describing such a model is given by
\begin{equation}
H=\int\dx\left(\frac{\mathrm{d}\hat{\Psi}^{\dagger}(x)}{\dx}\frac{\mathrm{d}\hat{\Psi}(x)}{\dx}+c\hat{\Psi}^{\dagger}(x)\hat{\Psi}^{\dagger}(x)\hat{\Psi}(x)\hat{\Psi}(x)\right),
\end{equation}
where $x\in[0,L]$ is the position coordinate and $c$ is the interaction strength. %\AS{[which is related to the interaction strength]}.

For our application, we generate $(Q,R)$ parametrisations of cMPS approximations for several bond dimensions of the Lieb-Liniger ground state for particular values of interaction strength $c$ by using the algorithm and implementation of v.~Hase \cite{cmps_groundstate}. This algorithm is an adaptation of the time-dependent variational principle for quantum lattices~\cite{2011PhRvL.107g0601H} to the continuous case (compare also Ref.\ \cite{2013PhRvL.111b0402D}). It relates to an imaginary time evolution that exponentially damps all excited components of an initial state vector $\ket{\Psi^{\left(d\right)}}$ (a cMPS with bond dimension $d$) with increasing imaginary time and produces the ground state eigenvector of a Hamiltonian $H$, by applying $\textrm{e}^{-\im Ht}$ with $t\in\im\mathbb{R}$ to $\ket{\Psi^{\left(d\right)}}$. The convergence of the energy of $\textrm{e}^{-\im Ht}\ket{\Psi^{\left(d\right)}}$ indicates the approach to the cMPS ansatz ground state vector, 
which we denote by $\ket{\Theta^{(d)}_{Q,R}}$, together with its characterising matrices $Q$ and $R$.
Several interesting structural properties of the state in the cMPS representation are revealed, signifying a symmetry in the model:
 degeneracies and a block structure of the matrix $M$. These features emerge in the integrable Lieb-Liniger case, and do not appear in Gaussian-sampled cMPS as described above.  
These features, which will be discussed more in detail in the following, appear regardless of the bond dimension and interaction strength used. Moreover, they do not depend on the algorithm used to obtain the ground state. 

%%%%%%%%%%%%%%%%%%%%%%%%%%%%%%%%%%%%%%%%%%%%%%%%%%%%%%%%%%%%%%%%%%%%%%%%%%%%%%

%\subsubsection{Source of correlation functions}
%%
%For our application, we use $Q$ and $R$ parametrisations of cMPS approximations for several bond dimensions of the ground state of the Lieb-Liniger Hamiltonian for particular values of interaction strength $c$ by using the algorithm and implementation of v.~Hase \cite{cmps_groundstate}. This algorithm is an adaptation of the time-dependent variational principle for quantum lattices~\cite{2011PhRvL.107g0601H} to the continuous case, see also Draxler et al.~\cite{2013PhRvL.111b0402D}. It relates to an imaginary time evolution that exponentially damps all excited components of an initial state vector $\ket{\Psi^{\left(d\right)}}$ (a cMPS with bond dimension $d$) with increasing imaginary time and produces the ground state eigenvector of a Hamiltonian $H$, by applying $\textrm{e}^{-\im Ht}$ with $t\in\im\mathbb{R}$ to $\ket{\Psi^{\left(d\right)}}$. The convergence of the energy of $\textrm{e}^{-\im Ht}\ket{\Psi^{\left(d\right)}}$ indicates the approach to the cMPS ansatz ground state, which we denote by $\ket{\Theta^{(d)}_{Q,R}}$, together with its characterising matrices $Q$ and $R$.

%%%%%%%%%%%%%%%%%%%%%%%%%%%%%%%%%%%%%%%%%%%%%%%%%%%%%%%%%%%%%%%%%%%%%%%%%%%%%%

\subsubsection{Degeneracies in the eigenvalue structure of $M$}

The topic of this section is to characterise the structure of the spectrum of $M$ by understanding the degeneracy structure $R$ in the exactly integrable case. 
In the case at hand, since all two-fold degenerate eigenvalues are equally spread into one of both blocks each, one is able to predict the spectrum of $R$ from $M$ even without reconstructing the second block.
In our simulations, it is seen that the eigenvalues of $Q$ and $R$ appear in $\lfloor d/2\rfloor $
pairs $\{ q_{j}^{[1]},q_{j}^{[2]}\} $
and $\{ r_{j}^{[1]},r_{j}^{[2]}\} $
with 
\begin{equation}
	q_{j}^{[1]}=\overline{q_{j}^{[2]}}+\im\chi,  ~~
		r_{j}^{[1]}=\overline{r_{j}^{[2]}}\textrm{e}^{\im\varphi},
\end{equation}	
respectively, for each pair $j$, with $\chi, \phi\in\mathbb{R}$ independent of $j$. If $d$ is odd, the two remaining unpaired
eigenvalues take the form $q=\hat{q}+\im\chi$ and $r=\hat{r}\textrm{e}^{\im\varphi}$,
respectively, with $\hat{q},\hat{r}\in\mathbb{R}$. We can simplify the structure by performing the
transformations
\begin{equation}
Q\mapsto Q-\im\chi\id1_{d}, ~~      R\mapsto R\textrm{e}^{-\im\varphi},\label{eq:applic:LL Q,R rot transl}
\end{equation}
which leave the transfer matrix
$T$ and all density-like $n$-point functions invariant.
This ensures that the pairs now consist of complex conjugates and the spectra of $Q$ and $R$ are closed under
complex conjugation, which we want to require for the further argument. 

Since the spectrum of $M$ by construction is the same as that of $\overline{R}\otimes R$
(up to a normalisation constant and each $\lambda\in\sigma\left(\overline{R}\otimes R\right)$
can be written as $\overline{r}_{j}\cdot r_{k}$ with certain $j,k= 1,\dots,d $,
the appearance of complex conjugate pairs in the spectrum of $R$
implies twofold degeneracies for the according eigenvalues in the
spectrum of $M$ as products of $R$ eigenvalues, especially 
\begin{equation}
	\overline{r_{j}^{[1]}}r_{k}^{[1]}=r_{j}^{[2]}\overline{r_{k}^{[2]}}=\overline{r_{k}^{[2]}}r_{j}^{[2]}.
\end{equation}
Not all eigenvalues are degenerate: $\overline{r_{j}^{[1]}}r_{j}^{[2]}$
and $\overline{r_{j}^{[2]}}r_{j}^{[1]}$ are
complex conjugates, but since $j=k$, there are no other combinations
that yield the same values. Assuming that $R$ does not contain any other degeneracies, $M$ will comprise $d$ non-degenerate eigenvalues and $d^2-d$ eigenvalues that are twofold degenerate each.

%The spectrum of $K$ features an interesting property as well. The
%\emph{differences} $\Delta\kappa_{j}:=\kappa_{j+1}-\kappa_{j}$ of
%the eigenvalues always appear in pairs so that $\Delta\kappa_{1}=\Delta\kappa_{d-1}$,\textrm{
%$\Delta\kappa_{2}=\Delta\kappa_{d-2}$} and so on.

%%%%%%%%%%%%%%%%%%%%%%%%%%%%%%%%%%%%%%%%%%%%%%%%%%%%%%%%%%%%%%%%%%%%%%%%%%%%%%

\subsubsection{Block structure}
Another structural observation we can make for the matrix $M$ of the ground state of the Lieb-Liniger Hamiltonian is the fact that it can be transformed to a block diagonal matrix. We do this by simply grouping vanishing and non-vanishing elements in $M$ and interchanging its rows and columns correspondingly, which amounts to a basis permutation. This way, we define the matrix $M^{\square}:=M_{1}\oplus M_{2}$, where $M_{1}$ and $M_{2}$ are block matrices and relate to the non-vanishing and vanishing residues of the cMPS. The block structure of $M^\square$ and the fact that $e^T$ is diagonal imply a block structure of their products, which carries over to the correlation functions, lets $M_2$ decouple completely, and hence disappear from the reconstruction.

We can see why all the residues corresponding to $M_{2}$ vanish for every $n$-point function in the following way. Let us assume we reordered $M$ and formed $M^{\square}$ by performing the basis permutations described above, and we consider a pole $\lambda_{l}$ of the cMPS. For an arbitrary $n$-point function, each residue which contains the index $l$ at least once can be written as
\begin{equation}
%\begin{split}
\rho_{k_{1},\dots,k_{j-1},l,k_{j+1},\dots,k_{n-1}}= M_{1,k_{1}}^{\square}\dots M_{k_{j-1},l}^{\square} M_{l,k_{j+1}}^{\square}\dots M_{k_{n-1},1}^{\square}
%\end{split}
\end{equation}
with $j= 2,\dots,n-2 $. %, see Eq.\ \eqref{eq:cMPS:expval: n point fun comp}.

We take $l\in\{ \zeta+1,\dots,d^{2}\} $, where $\zeta$ is the dimension of $M_1$, i.e., $\lambda_l$ corresponds to a pole associated with $M_2$. In this situation, we note that two things can happen. Either $k_{j+1}\leq\zeta$, and $M_{l,k_{j+1}}^{\square}=0$ since the entry is located in the lower left block of $M^{\square}$, which contains just zeros, and thus the entry vanishes. Or $k_{j+1}>\zeta$, and there exists an entry $M_{k_{m},k_{m+1}}^{\square}$ with $m>j$, $k_{m}>\zeta$, and $k_{m+1}\leq\zeta$ such that $M_{k_{m},k_{m+1}}^{\square}=0$. This has to eventually happen since the last entry in the residue expression is of the form $M_{k_{n-1},1}^{\square}$ and $1\leq\zeta$. Clearly, the residue vanishes again, and so does for the boundary indices $k_{1}=l$ or $k_{n-1}=l$.

%%%%%%%%%%%%%%%%%%%%%%%%%%%%%%%%%%%%%%%%%%%%%%%%%%%%%%%%%%%%%%%%%%%%%%%%%%%%%%

\subsubsection{Reconstruction}

Because of the block structure of $M$, we conclude that there is no direct way of obtaining all poles of cMPS approximations of the Lieb-Liniger ground state from an $n$-point density-like correlation function. In this case, the $p$-number \cite{Wick_MPS}, which is defined as the minimum order for a $p$-point function of a cMPS to reveal all poles, is infinite.
There is a useful connection between the degeneracies in the spectrum of $M$ and its block structure for the Lieb-Liniger model. It turns out that all the non-degenerate eigenvalues are related to $M^{\square}$ entries in the first block, while the degenerate pairs are distributed such that always one eigenvalue is associated with the first block and the other with the second. This way, since only the first block contributes to any density-like correlation function, all degeneracies are \emph{effectively} lifted, and hence full reconstruction is possible. 
Since all eigenvalues of $M$ that appear in the vanishing second block also appear in the visible first block one can in principle determine the spectrum of $R$ even without full knowledge of $M$. The same holds for the spectrum of $Q$ since also $D-M$ has the same spectral properties. For reconstructing \emph{both} $R$ and $Q$ in the corresponding gauge, however, our procedure requires full knowledge of $M$.
But again, note that for full reconstruction of the density-like correlation functions, this full knowledge is not required here.

This structure disappears if integrability is broken, and hence in a neighbourhood around the
(cMPS approximation of the) Lieb-Liniger ground state. Imaginary time evolution gives us a notion 
of distance to the limit of the approximation process, as we can, e.g., observe convergence of matrix entries along imaginary time paths. The block structure and degeneracy become more clearly defined the closer one gets to the limit point. Ultimately, at the limit point of the imaginary time evolution, the degeneracies and block structure of $M$ will prevent our methods to recover a full cMPS description in terms of matrices $Q$ and $R$ of the system. On the other hand, for each state along such a path, we can in principle apply our reconstruction method. The closer we get, the better all characteristic parameters can be reconstructed although the more ill-conditioned the problem becomes. A reconstruction of the $n$-point functions of arbitrary order is still possible, as it is based on the observable blocks of the matrices $D$ and $M$ alone and determining these quantities is in principle possible. Since the second block does not contribute to \emph{any} $n$-point function, the applicability of ``Wick's theorem'' for (continuous) matrix-product states is maintained even in this case and we still can successfully predict higher order from lower order correlation functions.

%%%%%%%%%%%%%%%%%%%%%%%%%%%%%%%%%%%%%%%%%%%%%%%%%%%%%%%%%%%%%%%%%%%%%%%%%%%%%%
%%%%%%%%%%%%%%%%%%%%%%%%%%%%%%%%%%%%%%%%%%%%%%%%%%%%%%%%%%%%%%%%%%%%%%%%%%%%%%
%%%%%%%%%%%%%%%%%%%%%%%%%%%%%%%%%%%%%%%%%%%%%%%%%%%%%%%%%%%%%%%%%%%%%%%%%%%%%%

\section{Summary and outlook}\label{sec:summary}

In this work, we have introduced the concept of quantum field tomography. In spite of the inherent difficulties of attempting to reconstruct a continuous system, i.e., a system with infinite degrees of freedom, we have shown that this task can be done when only a relevant class of naturally occurring states is considered. This is physically well motivated since one expects naturally appearing states not to be of the most general form but restricted to a smaller class of states. This is clearly the case in physical applications in which, for example, matrix product states have been shown to be a very successful model to describe correlations and dynamics. Here, we concentrated on developing tomographic tools for one-dimensional continuous many-body systems or quantum fields. 

For this purpose, we employed the continuous generalisation of the MPS variational class of states: the continuous matrix product state formalism. Based on this formalism and the predicted structure of the relevant data, i.e., the correlation functions, we developed a procedure to extract a best fit cMPS using state of the art statistical estimation tools. In this way, we are able to deliver a working and readily applicable tool to study this type of systems. The procedure we offer can indeed be seen as the natural way to think of efficient quantum field tomography. 
This does not mean, however, that for
tasks of direct estimation of fidelities and properties of states, alternative methods may not be advisable. The machinery here aims at reconstructing the states as such.

Formally, we have used the cMPS framework to describe the structure of correlation functions that can in principle be measured in experiments. Having identified this basic structure, we defined the tools needed to extract the pertinent information from the data. For this purpose, we employed the matrix pencil method as a viable way to determine the variational parameters of the cMPS from a correlation function. We showed that one can successfully extract a cMPS description of a system in principle for arbitrary bond dimensions. However, for noisy signals, one is in general limited to lower bond dimension approximations. Generally,
this approach is applicable to states with low entanglement, similarly to matrix-product states approximating states that satisfy an area law for suitable Renyi entanglement entropies. In the discrete case,
the connection of having ``low entanglement'' and being approximable with a matrix product state of low bond dimension has been fully rigorously fleshed out already 
\cite{SchuchApprox,EisertAreaLaws}. 
In the continuous case,
this connection is surely equally plausible, but is awaiting a similar fully rigorous treatment.

Moreover, we have given an in-depth study of the applicability of the reconstruction tools and their robustness for different noise models. Extensive numerical simulations were employed which provide at least empirical confidence of the performance of the reconstruction tools. We found that for the cases studied in this work, our methods are reasonably robust to noise when searching for low bond dimension cMPS estimates. 

It is important to note that the methods developed in this work are likewise readily applicable to the translationally invariant discrete MPS case. Since  in reality one deals with discrete (sampled) data even if the system is continuous in nature, all the methods developed here carry to the discrete case of matrix product states, reflecting a finite lattice spacing,
with minimal modifications. Furthermore, there is evidence that the approach taken here reveals insight into the structure of the underlying model as such and can detect signatures of 
integrability.

The novel methods proposed in this work open a new avenue to explore continuous systems of many particles in both equilibrium and non-equilibrium. 
It constitutes a step towards assessing strongly correlated models with a topographic mindset, without having to make a model of the system in the first place: Instead, one 
asks what the state is that is most compatible with the data found. This is a most healthy mindset specifically in the context of emergent quantum technologies, where one aims
at assessing the state of a quantum system without making too strong assumptions in the first place. In quantum information science, quantum state tomography is already a 
pillar on which the field rests, a technique routinely applied in most experiments. The present work opens up perspectives to think of quantum field tomography of strongly correlated
quantum systems, as they feature in dynamical quantum simulators. Specifically in this context, the tools presented here can be used for partial benchmarking of 
analog quantum simulators. To fully explore the potential of such an approach to study many-body systems out of equilibrium constitutes a 
truly exciting perspective.

\section{Acknowledgements}
This work was supported by the  the BMBF, the EU (RAQUEL, COST, SIQS, AQUS), the Dahlem Research School, 
and the ERC (TAQ). We thank T.\ J.\ Osborne and M.\ Friesdorf for discussions and M.\ von Hase for 
providing
the code using the time-dependent variational principle.
\bibliographystyle{acm}
%\bibliography{cmps}

\end{document}